\documentstyle[12pt,amsfonts]{article}
\parskip=\medskipamount

\newcommand {\cD}{{\cal D}}

\newcommand {\cF}{{\cal F}}

\newcommand {\cH}{{\cal H}}

\newcommand {\cM}{{\cal M}}
\newcommand {\cN}{{\cal N}}

\newcommand {\cQ}{{\cal Q}}
\newcommand {\cR}{{\cal R}}

\newcommand {\cV}{{\cal V}}
\newcommand {\cW}{{\cal W}}

\newcommand {\cY}{{\cal Y}}

\newcommand{\bC}{{\bf C}}

\newcommand{\bF}{{\bf F}}

\newcommand{\bN}{{\bf N}}

\newcommand{\bP}{{\bf P}}
\newcommand{\bQ}{{\bf Q}}
\newcommand{\bR}{{\bf R}}

\newcommand{\bV}{{\bf V}}
\newcommand{\bW}{{\bf W}}
\newcommand{\bX}{{\bf X}}

\newcommand{\bZ}{{\bf Z}}
\def\a{\alpha}
\def \bi{\bibitem}

\def\b{\beta}
\def\c{\chi}
\def\d{\delta}

\def\f{\phi}
\def\g{\gamma}
\def\G{\Gamma}

\def\j{\psi}

\def\l{\lambda}
\def\m{\mu}
\def\n{\nu}
\def\o{\omega}
\def\p{\pi}
\def\q{\theta}
\def\r{\rho}
\def\s{\sigma}

\def\x{\xi}
\def\z{\zeta}
\def\D{\Delta}
\def\F{\Phi}
\def\J{\Psi}

\def\O{\Omega}

\def\S{\Sigma}
\def\U{\Upsilon}

\def\tr{{\rm tr}}
\newcommand{\ad}{{\dot{\alpha}}}                           %new
\newcommand{\bd}{{\dot{\beta}}}                            %new
\newcommand{\ve}{\varepsilon}                            %new
\newcommand{\pa}{\partial}                           %new
\newcommand{\hf}{\frac12}
\newcommand{\vf}{\varphi}
\newcommand{\sect}[1]{\setcounter{equation}{0}\section{#1}}

\newcommand{\be}{\begin{equation}}
\newcommand{\ee}{\end{equation}}
\newcommand{\bea}{\begin{eqnarray}}
\newcommand{\eea}{\end{eqnarray}}
\newcommand{\non}{\nonumber}

\newcommand{\au}{\underline{a}}
\newcommand{\bu}{\underline{b}}
\newcommand{\cu}{\underline{c}}
\newcommand{\du}{\underline{d}}
\begin{document}
%%%%%%%%%%%%%%%%%%%%%%%%%%
%%%%%%%%%%%%%%%%

\begin{titlepage}
\thispagestyle{empty}

\begin{flushright}
hep-th/0210007 \\
AEI-2002-077\\
CERN-TH/2002-253\\
October, 2002
\end{flushright}
\vspace{5mm}

\begin{center}
{\Large \bf
Low energy dynamics from deformed
conformal symmetry in quantum 4D N = 2 SCFTs }
\end{center}
%\vspace{3mm}

\begin{center}
{\large S. M. Kuzenko,${}^\dagger$ I. N. McArthur${}^\dagger$
and S. Theisen${}^\ddagger$}\\
\vspace{2mm}

${}^\dagger$\footnotesize{
{\it
School of Physics, The University of Western Australia\\
Crawley, W.A. 6009. Australia}
} \\
{\tt  kuzenko@cyllene.uwa.edu.au},~
{\tt mcarthur@physics.uwa.edu.au}\\
\vspace{2mm}

${}^\ddagger$\footnotesize{
{\it
Max-Planck-Institut f\"ur Gravitationsphysik,
Albert-Einstein-Institut \\
Am M\"uhlenberg 1, D-14476 Golm, Germany }\footnote{permanent address}\\
and\\
{\it CERN Theory Division, CH-1211 Geneva 23, Switzerland}
} \\
{\tt theisen@aei-potsdam.mpg.de}
\end{center}
\vspace{5mm}

\begin{abstract}
\baselineskip=14pt
We determine the one-loop deformation of the conformal symmetry 
of a  general ${\cN}=2$ superconformally invariant Yang-Mills theory.
The deformation is computed for several explicit examples which have
a realization as world-volume theories on a stack of D3 branes.
These include (i) $\cN=4$ SYM with gauge groups $SU(N)$, $USp(2N)$ 
and $SO(N)$; (ii) $USp(2N)$ gauge theory with one hypermultiplet 
in the traceless antisymmetric representation and four hypermultiplets 
in the fundamental; (iii) quiver gauge theory with 
gauge group $SU(N) \times SU(N)$ and two hypermultiplets
in the bifundamental representations $(\bN, {\bar \bN} )$ 
and $({\bar \bN}, \bN)$. The existence of quantum corrections 
to the conformal transformations imposes restrictions 
on the effective action which we study on a subset of the Coulomb 
branch corresponding to the separation of one brane from the stack.
In the $\cN=4$ case, the one-loop corrected transformations provide 
a realization of the conformal algebra; this deformation is shown
to be one-loop exact. For the other two models, higher-loop corrections
are necessary to close the algebra. Requiring closure, 
we infer the two-loop conformal deformation.
\end{abstract}

\vfill
\end{titlepage}

\newpage
\setcounter{page}{1}

\renewcommand{\thefootnote}{\arabic{footnote}}
\setcounter{footnote}{0}
%%%%%%%%%%%%%%%%%%%%%%%%%%%
\sect{Introduction and summary}

Four-dimensional conformal field theories have attracted
much attention recently, mainly due to the possibility to study them
via a dual supergravity description; for a review, see \cite{AGMOO}.
This was first proposed for maximally supersymmetric
${\cal N}=4$ Yang-Mills theory \cite{Maldacena},
which has long been the prime example of a four-dimensional
conformally invariant quantum field theory. Here the
dual supergravity theory is type IIB supergravity compactified
on $AdS_5\times S^5$, which is the near-horizon geometry of a
stack of D3 branes. One simple argument in favour of the
gauge theory -- gravity correspondence is provided by comparison of
their symmetries.
The isometry group $SU(2,2)\times SO(6)$ of $AdS_5\times S^5$
coincides with the conformal group of four-dimensional Minkowski space
and the ${\cal R}$-symmetry group of ${\cal N}=4$ SYM theory.
This agreement can be extended to the full supergroup $SU(2,2|4),$
of which the above is the bosonic subgroup. From this comparison of
symmetries it is clear that it is the $AdS_5$ factor of the compactification
which is responsible for the conformal invariance of the dual
gauge theory. The fact that the beta-function of the field theory
vanishes is reflected by the constancy of the type IIB dilaton.
Subsequently, many generalizations of the original
proposal have been considered and dual supergravity descriptions
of conformal and confining gauge theories have been constructed
with various gauge groups and number of supercharges;
see, e.g. \cite{confining}.

In this paper we address the question to what extent one can infer the
geometry of the dual gravity background -- if any --
from the conformal field theory.
As to the latter, we restrict our attention to
conformally invariant supersymmetric Yang-Mills theories.
At the classical level they are invariant under conformal
transformations, $\d_{\rm c} \f^I$,
with respect to which the fields $\f^I$
transform in the standard way,
specified by their tensorial structure and their conformal weight.
One may then ask whether this symmetry is still manifest
(i.e. has the same functional form) in the effective action.
The issue here is that quantization requires gauge fixing
and the latter can be shown to necessarily break
(part of) the conformal symmetry (see \cite{FP1,FP2} and references
therein).
The effective action will thus not be
invariant under the same conformal transformations which were a symmetry
of the classical action. The change in the gauge fixing condition under
conformal transformations can be undone in the path integral
by a compensating field-dependent gauge transformation.
The invariance of the path integral
under combined conformal and gauge transformations leads to
modified conformal Ward identities for the effective action.
In other words, the effective action
is invariant under deformed conformal transformations,
\be
\d \f^I \, \frac{ \d \G [\f]}{\d \f^I} = 0~, \qquad
\d \f^I = \d_{\rm c} \f^I
+ \sum_{L=1}^{\infty} \hbar^L \,
\d_{(L)} \f^I~,
\label{multi-loop-deform}
\ee
and this deformation can, in principle, be computed at each loop order.
Of course, the deformation depends only on the parameters of those
transformations which do not leave the gauge fixing condition
invariant, and these transformations are the special conformal boosts.
Having obtained the deformed conformal transformations, this imposes
severe restrictions on the general form of the effective action.
Indeed, if the effective action were invariant under the classical
conformal transformations, each term in its loop expansion,
\be
\G [\f] = S[\f]
+ \sum_{L=1}^{\infty} \hbar^L \,
\G_{(L)} [\f]~,
\ee
would be conformally invariant. The deformed conformal
transformations, however, mix different orders in the loop expansion
of $\G[\f]$. In particular, it turns out that even
the one-loop deformation, $\d_{(1)} \f^I$, contains some
nontrivial information about the multi-loop structure
of the effective action.

If we now consider the field theory as living on a D3
brane in a ten-dimensional space-time, the deviation of the
position of the brane from a chosen reference position is parametrized
by the vacuum expectation values of massless scalar fields.
For the field theory this corresponds to going to the
Coulomb branch of the vacuum manifold.
If the ambient geometry is non-trivial, the effective action,
which is constrained by requiring it to be invariant under deformed
transformations, should provide information about this geometry.
Analysis of the one-loop deformed conformal symmetry on the
Coulomb branch of $\cN=4$ $SU(N)$ super Yang-Mills theory
has  been carried out in \cite{JKY,KM1,KM2}.

When constructing SYM theories and their quantum-deformed
symmetry properties we must, first of all, ensure
that the conformal symmetry survives quantization. A necessary
condition for this is the vanishing of the beta-function, which can be
arranged by an appropriate choice of matter fields. In the
${\cal N}=4$ SYM theory there is no freedom and the theory
is completely specified by a choice of gauge group. For ${\cal N}=2$
the beta-function is one-loop exact and conformal invariance
imposes a single condition on the second Casimir invariants of the
matter hypermultiplets \cite{HSW}.
${}$For ${\cal N}=1$ the situation is more
complicated and one generally finds lines of superconformal fixed
points in a higher-dimensional moduli space of vacua
\cite{LS} (this paper also includes an extensive list of references
on finite $\cN=1$ theories).
In the present paper we concentrate on conformally invariant SYM theories
with eight supercharges, i.e. ${\cal N}=2$.
For these we will explicitly determine the one-loop deformation of the
conformal transformation properties of the fields.

The remainder of the paper is organized as follows. In the next section
we first recall some well-known facts about ${\cal N}=2$ SYM, including
its background field quantization. Following \cite{KM1,KM2}
we collect all necessary formulas which are needed to compute the
deformed conformal transformation. In sect.~3 we do this, in
full generality, for an arbitrary ${\cal N}=2$ SYM at one loop order.
Sect.~4 is devoted to a discussion of the
conformal deformation on the Coulomb branch with vanishing
hypermultiplets.
Here a non-renormalization theorem guarantees that the
one-loop deformation is exact.\footnote{This holds in 't Hooft
gauge. Changing from this gauge to, say, $R_\x$ gauge
is equivalent to a nonlocal field redefinition in
the effective action, which leads to a restructure of the loop
expansion accompanied by a modification of 
the functional form of symmetries \cite{KM2}.}
One test of this is that the
deformed algebra closes without the need to add higher loop
contributions.
We then discuss to what extent invariance of the
deformed transformation fixes the form of the effective action
on the Coulomb branch.
In particular, if we choose the background fields
to correspond to moving one D3 brane away from a stack
of $N$ D3 branes, we are probing the geometry produced by them.
This geometry should then determine the general structure of the
effective action.

In sect.~5 we specialize to  ${\cal N}=4$ theories,
which are equivalent to
${\cal N}=2$ SYM theories with one hypermultiplet in the adjoint
representation of the gauge group. The brane constructions of these theories
are known: in the case of a stack of $N$ D3 branes, one obtains the
gauge group $SU(N)$; this is the case considered by
Maldacena \cite{Maldacena}.
If one puts $N$ D3 branes on top of an orientifold D3 brane, one obtains,
depending on the choice of the orientifold projection, either
$SO(2N)$ or $USp(2N)$ gauge groups
\cite{Witten} (in the $SO$ case $N$ can be a
half-integer). The near-horizon geometry of this brane configuration
is $AdS_5\times \bR \bP^5$. We compute the one-loop exact
deformed conformal transformation for all these cases.

In sects.~6 and 7 we consider two examples of ${\cal N}=2$ theories.
First, in sect. 6, we study the field theory which one obtains by
placing $N$ D3 branes on top of four D7 branes which are
coincident with an $O7$ plane. This gives a $USp(2N)$ gauge group
and one hypermultiplet in the traceless antisymmetric representation
and four hypermultiplets in the fundamental
representation \cite{ASTY,DLS,APTY}.
Here the near-horizon geometry is $AdS_5 \times S^5 / \bZ_2$
where the $\bZ_2$ action has a fixed $S^3 \subset S^5$
\cite{Fayyazuddin}.
Other examples of superconformal ${\cal N}=2$ theories are obtained
by considering stacks of $N$ D3 branes at an $\bR^4/\bZ_k$
singularity. This leads to so-called quiver gauge theories
\cite{DM,JM} with
gauge group $SU(N)^k$ and matter in the bifundamental representation
of adjacent gauge groups in the quiver diagram, which is a
$k$-gon in this simple case. The near-horizon geometry of this
brane configuration is $S^5/\bZ_k$ where the $\bZ_k$ action leaves
a $S^1\subset S^5$ invariant \cite{KS}.
In sect.~7 we consider the simplest case
of such theories, namely $k=2$. The generalization to other $k$'s
is straightforward.

While the one-loop deformation in ${\cal N}=4$ theories and on
the pure Coulomb branch of ${\cal N}=2$ theories is exact,
this is no longer the case for the mixed branches of
${\cal N}=2$ theories. One consequence of this is that the
conformal algebra does not close as long as the 
gauge group has finite rank. 
Requiring closure one can, on the
other hand, infer the form of the higher loop corrections to the
conformal transformation of the various fields.
We will determine them explicitly at two loop order.

\sect{\mbox{$\cN = 2$} SCFTs}

We consider a four-dimensional $\cN=2$ superconformal field theory,
which describes the coupling of an $\cN=2$ vector supermultiplet to
a massless hypermultiplet in a (possibly reducible)
representation R of the gauge group $G$.
The finiteness condition (which coincides with the requirement
of absence of one-loop divergences)  \cite{HSW}
can be given in the form
\be
{\rm tr}_{\rm Ad}\, W^2 = {\rm tr}_{\rm R}\, W^2~,
\label{finiteness}
\ee
where the subscript ``Ad'' denotes the adjoint representation,
and $W$ is an arbitrary complex scalar field taking its values in the
Lie algebra of the gauge group, $W= W_{\au} T_{\au}$,
with $T_{\au} =(T_{\au})^\dagger$ the gauge group generators.

The $\cN=2$ vector multiplet is composed of a gauge field $V_m$,
adjoint scalars $W$ and $\bar W = W^\dagger$,  and adjoint spinors
$\l^i_\a$ and ${\bar \l}_{\ad i} = (\l^i_\a )^\dagger$, where
$i=1,2$.
The hypermultiplet is
described by R-representation scalars and spinors
$( Q_i, \j_\a, {\bar \m}_\ad ) $
and their conjugates $( {\bar Q}^i, {\bar \j}_\ad, \m_\a )$,
where ${\bar Q}^i= (Q_i)^\dag $.
The Lagrangian (with $g$ the coupling constant)
is\footnote{Here
and in the following,
lower case Latin letters from the middle of the alphabet,
$i, j, k$,
are used to denote indices of the automorphism group of the
$\cN=2$ supersymmetry algebra, or the $\cR$-symmetry group
$SU(2)_\cR$. Such  indices are raised and lowered
by antisymmetric tensors $\ve^{ij}$ and $\ve_{ij}$, with
$\ve^{12}=\ve_{21}=1$, in the standard way:
$Q^i = \ve^{ij}\, Q_j$,  $Q_i = \ve_{ij}\, Q^j$,
such that $(Q_i)^\dag = {\bar Q}^i$,
$(Q^i)^\dag = -{\bar Q}_i$.}
\bea
g^2\, L &=& - {\rm tr}_{\rm F} \,\Big\{ {1 \over 4} F^{mn} F_{mn}
 + D^m {\bar W} D_m W  +\hf [{\bar W}, W]^2
 \non \\
&&
+{\rm i} \,\l^i \s^m D_m {\bar \l}_i
-\frac{\rm i}{ \sqrt{2} }\, \l^i [{\bar W}, \l_i ]
+ \frac{\rm i}{ \sqrt{2} }\, {\bar \l}_i [ W, {\bar \l}^i ]
 \Big\} \non \\
 && -D^m {\bar Q}^i D_m Q_i - {\bar Q}^i \{ {\bar W}, W \} Q_i
-{\bar Q}^i T_{\au} Q_j \, {\bar Q}^j T_{\au} Q_i
+\hf ({\bar Q}^i T_{\au} Q_i )^2  \non \\
&& -{\rm i} \,\m  \s^m D_m {\bar \m}
-{\rm i} \,{\bar \j} \tilde{\s}^m D_m \j
- \sqrt{2} {\rm i}\, \m W \j
+  \sqrt{2} {\rm i}\, {\bar \j} {\bar W} {\bar \m} \non \\
&&+ \frac{\rm i}{ \sqrt{2} }
( {\bar Q}^i \l_i \j + {\bar \j} {\bar \l}^i Q_i )
+ \frac{\rm i}{ \sqrt{2} }
( {\bar Q}^i {\bar \l}_i {\bar \m} - \m \l^i Q_i )~,
\label{classact}
\eea
where $D_m = \pa_m +{\rm i}\, V_m$, and the generators
are normalized such that
${\rm tr}_{\rm F} (T_{\au} T_{\bu}) = \d_{\underline{ab}}$
in the fundamental representation.
The $Q{\bar Q}$ self-interaction occurs after elimination
of the auxiliary triplet, $X_{ij} = X_{(ij)}$,
which belongs to the off-shell $\cN=2$  vector multiplet.
This self-interaction can be rewritten
as\footnote{By giving up manifest $SU(2)_\cR$ invariance,
this potential can be brought to a more familiar form.
Defining $Q_1 \equiv Q$ and $Q_2 \equiv \tilde{Q}^\dagger$,
and thus ${\bar Q}^1 = Q^\dagger$ and ${\bar Q}^2 = \tilde{Q}$,
one obtains ${\bar Q}^{(i} T_{\au} Q^{j)}
{\bar Q}_{(i} T_{\au} Q_{j)}
= -2 |\tilde{Q}T_{\au} Q|^2 - \hf D_{\au} D_{\au}$,
where $D_{\au} = Q^\dagger T_{\au} Q
- \tilde{Q} T_{\au} \tilde{Q}^\dagger$.}
\be
-{\bar Q}^i T_{\au} Q_j \, {\bar Q}^j T_{\au} Q_i
+\hf ({\bar Q}^i T_{\au} Q_i )^2
= {\bar Q}^{(i} T_{\au} Q^{j)} \,
{\bar Q}_{(i} T_{\au} Q_{j)}~.
\ee

The model (\ref{classact}) admits a manifestly $\cN=2$
supersymmetric formulation with finitely many auxiliaries
in terms of constrained superfields \cite{GSW,Sohnius}
(in this approach, the hypermultiplet possesses an intrinsic off-shell
central charge). It can also be formulated in terms of unconstrained
superfields, which involve infinitely many auxiliary fields,
in harmonic superspace \cite{GIOS}. In both superfield realizations,
the $\cR$-symmetry $SU(2)_\cR $ is manifest.
The superconformal symmetry $SU(2,2|2)$ is manifest in the harmonic
superspace approach.

The moduli space of vacua of the theory under consideration
is specified by the following conditions:
\be
[{\bar \cW}, \cW ] = 0~, \qquad \cW \cQ_i = 0~, \qquad
{\bar \cQ}^{(i} T_{\au} \cQ^{j)} = 0~,
\label{flat}
\ee
with ${\bar \cW} \cQ_i = 0$ a consequence of the first and second
conditions. The solutions to the vacuum equations
(\ref{flat}) can be classified according to the phase
of the gauge theory they give rise to. In the pure Coulomb
phase the rank of the gauge group is unreduced:
generically it corresponds to
$\cQ_i=0$ and $\cW \neq 0$ and unbroken gauge group 
$U(1)^{{\rm rank}(G)}$. 
In the (pure) Higgs phase the gauge
symmetry is completely broken; there are no massless gauge bosons.
This requires $\cQ_i\neq 0$.
In the mixed phases there are some massless gauge bosons but the
rank of the gauge group is reduced.

At tree level at energies below the symmetry breaking scale,  
we have free field massless dynamics
if the $\cN=2$ vector multiplet $( \cV_m, \cW, \ldots )$
and the hypermultiplet $( \cQ_i, \ldots )$ are
aligned along a particular  direction in the moduli space of vacua.
At the quantum level, however,
exchanges of virtual massive particles produce corrections to
the action of the massless fields. The aim of this work is
to determine restrictions on the structure of the low energy
effective action which are implied by quantum conformal invariance
of the theory.

We quantize the $\cN=2$ SYM theory (\ref{classact})
in the framework of the background field method
(see \cite{DeWitt67,tH} and references therein),
by splitting the dynamical variables
$\F^I = (V_m, W, Q_i, \ldots)$ into the sum of {\it background}
fields\footnote{For most of this section, the background fields
are completely arbitrary. After eq. (\ref{FR2}),
they will be taken to be
aligned along a particular  direction in the moduli space of vacua.}
$\f^I = (\cV_m , \cW, \cQ_i, \ldots)$ and {\it quantum}
fields $\vf^I = (v_m, w, q_i, \ldots)$.
The classical action, $S[\F]$, is invariant under standard
Yang-Mills gauge transformations
\be
\d V_m = - D_m \z
~,
\quad \d W = {\rm i}\, [\z, W]~, \quad
\d Q_i = {\rm i}\, \z Q_i
~,  \quad \ldots
\ee
which, in a condensed notation \cite{DeWitt67},
read
\be
\d \F^I = R^I{}_{\au} [\F] \,  \z_{\au}~,
\ee
with $R^I{}_{\au} [\F]$ the gauge generators and
$ \z_{\au}$ infinitesimal gauge parameters.
Upon background quantum splitting, the action
$S[\f + \vf] $ is invariant under
{\it background} gauge transformations
\be
\d \f^I = R^I{}_{\au} [\f] \, \z_{\au}~, \qquad
\d \vf^I = R^I{}_{{\au},J} \, \vf^J \,\z_{\au}~,
\label{back-trans}
\ee
and {\it quantum} gauge transformations
\be
\d \f^I = 0~, \qquad
\d \vf^I = R^I{}_{\au} [\f +\vf] \,\z_{\au}~.
\ee
The background field quantization procedure
consists of fixing the quantum gauge freedom
while  keeping  the background gauge invariance intact
by means of  background  covariant gauge conditions.
The effective action is given by the sum of all 1PI Feynman graphs
which are vacuum with respect to the quantum fields.

The quantum gauge freedom will be fixed by choosing the following
background covariant gauge conditions
(often called 't Hooft gauge):
\be
\c_{\au} = \cD \cdot v_{\au} + {\rm i} \,{\bar w} T_{\au} \cW
-{\rm i} \,{\bar \cW} T_{\au} w
+ {\rm i} \,{\bar q}^i T_{\au} \cQ_i
- {\rm i}\, {\bar \cQ}^i T_{\au} q_i ~,
\label{thooft}
\ee
and the gauge fixing functional\footnote{A note on notation:
while e.g. in (\ref{flat}) $\cW$ is matrix valued, it is an
(adjoint) vector in (\ref{thooft}). We will freely switch from one to the
other form to simplify expressions.}
\be
S_{\rm GF} [\c] = -\frac{1}{ 2g^2} \int {\rm d}^4 x
 \, \c^2 ~.
\ee
In eq. (\ref{thooft}), $\cD$ denotes the background gauge covariant
derivative, $\cD_m = \pa_m +{\rm i}\,\cV_m$.
Introducing
\bea
Y = \left(
\begin{array}{c}
W \\
Q_i
\end{array}
\right) =\cY + y~, \qquad
{\bar Y}  = ( {\bar W} , ~~ {\bar Q}^i ) = {\bar \cY} +{\bar y}~,
\eea
the gauge conditions can be rewritten in the
abbreviated form
\be
\c_{\au} = \cD \cdot v_{\au} + {\rm i} \,{\bar y} T_{\au} \cY
-{\rm i} \,{\bar \cY} T_{\au} y~.
\ee
Under the quantum gauge transformations,
$\c=\c[\vf,\f]$ changes as
\be
\d \c_{\au} = - (\cD^m D_m \z)_{\au}
+ {\bar \cY} T_{\au} \z (y + \cY)
+ ({\bar y} + {\bar \cY}) \z T_{\au} \cY
\equiv (\D \z)_{\au}~,
\ee
with $\D= \D[\vf,\f]$ the Faddeev-Popov operator.
To define the effective action, $\G[\f]$,
let us introduce the generating functional of
connected quantum Green's functions, $ W[J,\f ]$,
\be
{\rm e}^{{\rm i} W[J, \f]} =
N \int \cD \vf \, {\rm Det}  (\D[\vf, \f]) \,
{\rm e}^{ {\rm i} ( S[\f +\vf ] + S_{\rm GF} [\c [\vf, \f ]]
+ J_I \vf^I)}~.
\label{gen-fun-quan}
\ee
Its Legendre transform,
\be
\G[\langle \vf \rangle, \f ]
= W[J , \f ] -  J_I \, \langle \vf^I \rangle~,
\qquad
\langle \vf^I \rangle = \frac{ \d}{ \d J_I} W[ J, \f]~,
\ee
is related to the
effective action  $\G [\f]$ as follows:
$\G[\f] =\G[ \langle \vf \rangle =0, \f]$.
In other words, $\G [\f]$ coincides with
$W[ J,\f ]$ at its stationary point $J = J[\f]$
defined by $\d W[ J ,\f] /\d J = 0$.
By construction, $\G [\f]$ is invariant
under background gauge transformations.

The theory under consideration is conformally invariant,
both at the classical and quantum levels.
The classical action does not change under standard linear
conformal transformations of the fields
(these transformations differ in sign from those
adopted in \cite{KM1,KM2}):
\bea
\d_{\rm c} V_m  &=&  \x V_m +\o_m{}^n V_n + \s V_m~, \quad
\d_{\rm c} W =  \x W +\s W~, \quad
\d_{\rm c} Q_i =  \x Q_i +\s Q_i~, \non \\
\d_{\rm c} \J &=& \x \J + \hf\, \o^{mn} L_{mn}\J
+ {3\over 2} \, \s \J~,
\label{conf}
\eea
where $\J$ denotes any spinor field, $L_{mn}$ the Lorentz
generators in the spinor representation, and
$\x= \x^m \pa_m$  an arbitrary conformal Killing vector field,
\be
\pa_m \x_n + \pa_n \x_m = 2\eta_{mn}\, \s~, \qquad
\s \equiv \frac{1}{4}\pa_m \x^m~, \qquad
\o_{mn} \equiv \hf (\pa_m \x_n -  \pa_n \x_m)~.
\label{cKe}
\ee
At the quantum level, conformal invariance is governed
by the Ward identity \cite{KM1,KM2}
\be
\d_{\rm c} \f^I [\f] \, \frac{\d \G [\f]}{\d \f^I}
= \langle R^I{}_{\au} [\f +\vf] \,
(\D^{-1} [\vf,\f] \, \r [\vf] )_{\au} \rangle \;
\frac{\d \G[\langle \vf \rangle, \f]}
{\d \langle \vf^I  \rangle } \Big|_{ \langle \vf \rangle =0} ~,
\label{def-final}
\ee
where $\r [\vf]$ denotes the inhomogeneous
term\footnote{The general solution to
the conformal Killing equation (\ref{cKe}) is
$\x^m = a^m + \l x^m + K^m{}_n x^n +b^m x^2 - 2x^m (b \cdot x)$,
where the parameters $a^m$ and $K^{mn}= -  K^{nm}$ generate Poincar\'e
transformations, $\l$ dilatations and $b^m$ special conformal boosts.
By definition, $\s = \l - 2 (b \cdot x)$, hence  $\pa_m \s = -2b_m$,
and therefore the right-hand side in (\ref{def-final}) is
non-vanishing for  the special conformal boosts only.}
in the conformal transformation of $\c[\vf, \f]$:
\be
\d_{\rm c} \c =   (\x +2 \s) \, \c + \r~, \qquad
\r = - 2  \, (\pa^m \s ) \, v_m~.
\label{gauge-cond-conf-tran}
\ee
In (\ref{def-final}), the symbol
$\langle ~~~ \rangle $ denotes
the quantum average in the presence of sources,
\be
\langle F[ \vf, \f  ] \rangle
~=~ {\rm e}^{-{\rm i} W[J, \f]} \,
N \int \cD \vf \, F[\vf, \f] \,{\rm Det}  (\D[\vf, \f]) \,
{\rm e}^{ {\rm i} ( S[\vf +\f ] + S_{\rm GF} [\c [\vf, \f ]]
+ J_I \vf^I)}~.
\ee
Eq. (\ref{def-final}) should be treated in conjunction with
the identity \cite{KM1,KM2}
\bea
\d \f^I \, \frac{\d \G[\f]}{\d \f^I}
&=& \Big\{ \d \f^I +
\langle R^I{}_{\au} [\f +\vf] \,
(\D^{-1} [\vf,\f] \, \d \c [\vf, \f])_{\au} \rangle
\Big\}\,
\frac{\d \G[\langle \vf \rangle, \f ]}
{\d \langle \vf^I  \rangle } \Big|_{ \langle \vf \rangle =0}~, \non \\
\d \c [\vf, \f] &\equiv& \c [\vf -\d \f, \f +\d \f] -\c [\vf, \f] ~,
\label{FR2}
\eea
with $\d \f^I$ an arbitrary variation of the background fields.
This identity allows one to express
the functional derivative
$\d \G[\langle \vf \rangle, \f ] / \d \langle \vf \rangle $
at $\langle \vf \rangle = 0$ via $\d \G[\f] / \d \f$.

It follows from eqs. (\ref{def-final}) and (\ref{FR2})
that the effective action $\G[\f]$ is invariant under
quantum corrected conformal transformations,
as described by eq. (\ref{multi-loop-deform}).
In this paper we will evaluate the one-loop quantum deformation,
$\d_{(1)} \f$, of conformal symmetry when the fields are
aligned along a particular  direction
in the moduli space of vacua.

In the above discussion, the fields $\f^I$ have been completely
arbitrary. From now on, the background $\cN=2$ vector multiplet
and hypermultiplet will be chosen to be aligned
along a fixed, but otherwise arbitrary,
direction in the moduli space of vacua;
in particular, their scalar fields should solve the equations
(\ref{flat}). At later stages in this work,
we will be forced to impose further restrictions on the fields,
of the form
\be
\cV_m = \bV_m (x) H~, \quad  \cW = \bW (x) H ~;
\qquad \cQ_i = \bQ_i (x)  \U ~,
\label{separ-of-variables}
\ee
corresponding to a separation of space-time
and internal variables.
Here $H$ is a fixed generator in the Cartan subalgebra,
and $\U$ a fixed vector in the R-representation space of the gauge group,
in which the hypermultiplet takes values, chosen so that
$H \U =0$ and ${\bar \U} T_{\au} \U =0$, cf. (\ref{flat}).
The freedom in the choice of $H$ and $\U$ can be reduced by
requiring the field configuration (\ref{separ-of-variables})
to be invariant under a maximal unbroken gauge subgroup.
Eq. (\ref{separ-of-variables}) defines a single  $U(1)$
vector multiplet and a single hypermultiplet which is neutral
with respect to the $U(1)$ gauge subgroup generated by $H$.
It worth noting that an Abelian vector field and six neutral
scalars in four space-time dimensions is what we need to describe
a (static gauge) D3 brane moving in a ten-dimensional space-time.

Consider background gauge transformations
(\ref{back-trans}) with $\z =\z (x) H$ which will be
referred to as $H$-gauge transformations.
They leave all background fields  (\ref{separ-of-variables}) unchanged,
except the Abelian gauge field $\bV_m$.
By construction, such transformations leave invariant
the gauge-fixed action
$S[\f +\vf ] + S_{\rm GF} [\c [\vf, \f ]]$ and, in fact,
each term in its Taylor expansion in powers of the quantum
fields, since the background gauge transformations of the quantum
fields are linear and homogeneous.

\sect{The one-loop deformation}

${}$For the purpose of loop calculations,
we  expand the action $S[\f +\vf]$
in powers of the quantum fields $\vf$ and
combine its quadratic part, $S_2$,
with the gauge fixing functional, $S_{\rm GF}$.
Modulo fermionic contributions, the quadratic action is
\bea
S_2 &+& S_{\rm GF} = \frac{1}{g^2} \int {\rm d}^4x \,
\Big\{  - \hf \, v^m \tilde {\D} v_m + {\rm i}\, v^m \cF_m{}^n  v_n
- {\bar w} \tilde {\D} w
\non \\
&+& {\bar q}^i \cD^m \cD_m q_i
- {\bar q}^i \{ {\bar \cW}, \cW \} q_i
+ ({\bar q}^i T_{\au} \cQ^j + {\bar \cQ}^j T_{\au} q^i ) \,
( {\bar q}_i T_{\au} \cQ_j + {\bar \cQ}_j T_{\au} q_i ) \non \\
&+& 2{\rm i}\,( v^m (\cD_m {\bar \cW} ) w
-{\bar w} (\cD_m \cW) v^m )
+ 2{\rm i}\,( {\bar q}^i v^m \cD_m \cQ_i
- (\cD_m {\bar \cQ}^i) v^m q_i )  \Big\}~,
\label{quadrat-action}
\eea
where $\tilde{\D}$ is the Faddeev-Popov operator at $\vf =0$,
\be
( \tilde {\D} \z )_{\au}  = - (\cD^m \cD_m \z
- \{ {\bar \cW}, \cW \} \z)_{\au}
+ {\bar \cQ}^i \{ T_{\au}, T_{\bu} \} \cQ_i \, \z_{\bu}
= - (\cD^m \cD_m \z )_{\au}
+ {\bar \cY} \{ T_{\au},  \z \}\cY~.
\label{tilde}
\ee
It is assumed in (\ref{quadrat-action}) that the background
$\cN=2$ vector multiplet and hypermultiplet
are aligned along a particular direction
in the moduli space of vacua such as to satisfy (\ref{flat}).
The action (\ref{quadrat-action}) determines
the background covariant propagators of quantum fields,
$<\vf^I (x) \,  \vf^J (x')>$, which are required
to evaluate $\d_{(L)} \f $.

The one-loop deformation of the conformal transformations of the
fields can be computed by minor modification of the method described
in \cite{KM1} for the case of $\cN=4$ SYM.
The conformal Ward identity (\ref{def-final}) can be rewritten in the form
\bea
0 = \d_{\rm c} \cV_{m{\au}}  \, \frac{\d \G [\f]}{\d \cV_{m{\au}}}
&-& 2 (\partial^n \sigma) \,
\langle (D_m \Delta^{-1}
\, v_n)_{\au}\rangle \;
\frac{\d \G[\langle \vf \rangle, \f]}
{\d \langle v_{m{\au}} \rangle } \Big|_{ \langle \vf \rangle =0}
\nonumber \\
+
\d_{\rm c} \cY  \, \frac{\d \G [\f]}{\d \cY}
&+& 2 \,{\rm i} \, (\partial^n \sigma) \,
\langle (\Delta^{-1}\,
v_n)_{\au}
\, (T_{{\au}} y+ T_{{\au}} \cY) \rangle \;
\frac{\d \G[\langle \vf \rangle, \f]}
{\d \langle y \rangle } \Big|_{ \langle \vf \rangle =0} \nonumber \\
+ \d_{\rm c} \bar{\cY}  \, \frac{\d \G [\f]}{\d \bar{\cY}}
&-& 2 \,{\rm i} \, (\partial^n \sigma) \,
\langle (\bar{y}T_{{\au}} + \bar{\cY} T_{{\au}} )
\,
(\Delta^{-1}
\,v_n)_{\au}
 \rangle \;
\frac{\d \G[\langle \vf \rangle, \f]}
{\d \langle {\bar y} \rangle } \Big|_{ \langle \vf \rangle =0}~.
\label{WI}
\eea
At the one-loop level,
\bea
-2 (\pa^n \s)\,\langle
(D_m \Delta^{-1}
\, v_n)_{\au}\rangle
&=&
 \left( \delta_m{}^n \delta_{{\au}{\bu}}
+   (\cD_m \tilde{\Delta}^{-1} \cD^n )_{{\au}{\bu}} \right) \, \d_{(1)}
\cV_{n{\bu}}  \nonumber \\
&-& {\rm i} \, ( \cD_m \tilde{\Delta}^{-1} )_{{\au}{\bu}} \left(
\bar{\cY} \, T_{{\bu}} \, \d_{(1)}
\cY - \d_{(1)} \bar{\cY} \, T_{{\bu}} \, \cY \right)
\label{varv}
\eea
and
\bea
&& 2 \,{\rm i} \, (\partial^n \sigma)
\langle ( \Delta^{-1}
\,v_n)_{\au}
\, (T_{{\au}} y + T_{{\au}} \cY) \rangle = \, \left(1 - T_{{\au}} \cY \,
(\tilde{\Delta}^{-1})_{{\au}{\bu}} \, \bar{\cY} T_{{\bu}} \right) \,
\d_{(1)} \cY
\nonumber
\\
\,\,\,\,\,&&+ T_{{\au}} \cY \, (\tilde{\Delta}^{-1})_{{\au}{\bu}} \,
\left(
\d_{(1)}
\bar{\cY}\,  T_{{\bu}} \, \cY\right)
- {\rm i} \, T_{{\au}} \cY \, (\tilde{\Delta}^{-1} \cD^m)_{{\au}{\bu}} \,
\d_{(1)}
\cV_{m{\bu}}~,
\label{vary}
\eea
where
\bea
\d_{(1)}
\cV_{m{\au}}  &=& -2 \, {\rm i} \, (\pa^n \s)\,
(T_{{\cu}} \,\tilde{\D}^{-1} )_{{\au}{\bu}} \,
<v_{ n {\bu} } (x) \,  v_{m{\cu}} (x')>|_{x' = x} ~, \label{vector1}
  \\
\d_{(1)}
\cY &=&  2\, {\rm i} \, (\pa^n \s)\,
(\tilde{\D}^{-1} )_{{\au}{\bu}} \,
<v_{n{\bu}} (x) \,  T_{{\au}} \,y (x')>|_{x' = x} ~  \label{scalar1}
\eea
and $\tilde{\Delta}$ is given by eq. (\ref{tilde}).
As in the $\cN=4$ case \cite{KM1}, substitution of (\ref{varv}) and
(\ref{vary})   into
(\ref{WI}) yields
the  appropriate one-loop versions of the combination
\be
\d \f^I +
\langle R^I{}_{{\au}} [\f +\vf] \,
(\D^{-1} [\vf,\f] \, \d \c [\vf, \f])_{{\au}} \rangle 
\label{olc}
\ee
appearing in the identity (\ref{FR2}) to allow conversion of derivatives
of the effective action with respect
to quantum fields into derivatives of the effective action with
respect to background fields. The
conformal Ward identity then takes the one-loop form
\bea
0 &=& \left( \d_{\rm c} \cV_{m{\au}} + \d_{(1)}
\cV_{m{\au}} \right) \,
\frac{\d \G [\f]}{\d \cV_{m{\au}}}
+ \left( \d_{\rm c} \cY  + \d_{(1)}
\cY
\right) \, \frac{\d \G [\f]}{\d \cY}
+ \left( \d_{\rm c} \bar{\cY}  + \d_{(1)}
\bar{\cY}
\right) \, \frac{\d \G [\f]}{\d \bar{\cY} }~.
\eea
The one-loop deformations of the conformal field
transformations  are therefore given by
(\ref{vector1}) and (\ref{scalar1}).
The right-hand sides in (\ref{vector1}) and (\ref{scalar1})
are nonlocal functionals of $\cV_m$ and $\cY,$
each of which can be represented
as a sum of infinitely many local terms
with increasing number of derivatives of the fields.
We are interested in evaluating $\d_{(1)} \cV_m$ and
$\d_{(1)} \cY$ to first order in the derivative expansion.

To first order in the derivative expansion,
\be
<v_{ m {\bu} }  \,  w_{\cu}  > =2g^2
\Big( \tilde{\D}^{-1} \, (\cD_m \cW) \, \tilde{\D}^{-1}
\Big)_{ {\bu} {\cu}}
=2g^2 \Big( \tilde{\D}^{-2} \, (\cD_m \cW) \Big)_{ {\bu} {\cu}} ~,
\ee
where the last relation follows from the identity
\be
(\cD_m \cW)_{ {\bu} {\cu} } \,
{\bar \cY} \{ T_{\cu}, T_{\du} \} \cY
= {\bar \cY} \{ T_{\bu}, T_{\cu} \} \cY \,
(\cD_m \cW)_{ {\cu} {\du} } ~,
\ee
which is valid since $ \cW \,\cY = (\cD_m \cW) \, \cY = 0$
due to (\ref{flat}). As a result, we get
\be
\d_{(1)} \cW_{\au} =  4{\rm i}\,  g^2\,  (\pa^m \s)\,
{\rm tr}_{\rm Ad} \,\Big(T_{\au}{} \, (\tilde{\D}^{-3}| ) \, \cD_m \cW \Big) ~
\label{scalar2}
\ee
where $\tilde{\D}^{-3}|$ denotes the kernel of $\tilde{\D}^{-3}$
at coincident space-time points.
Let us now turn to the vector field variation.
To first order in the derivative expansion, one
similarly gets
\be
<v_{ m {\au} } \, v_{n {\bu} }  >
= - {\rm i} g^2  (\tilde{\D}^{-1})_{ {\au} {\bu} }\,
\d_{mn} + 2g^2 (\tilde{\D}^{-2} \, \cF_{mn})_{ {\au} {\bu} }~.
\ee
When substituted into (\ref{vector1}), the first term in the vector
propagator yields a potentially divergent contribution. However, it
vanishes on symmetry grounds, as
\be
(T_{{\cu}} \, \tilde{\Delta}^{-2})_{{\au} {\cu}} = - {\rm i} \,
f_{{\cu}{\au}{\du}} \, (\tilde{\Delta}^{-2})_{{\du} {\cu}},
\ee
and the mass matrix in $\tilde{\Delta}$ is symmetric under interchange of
${\cu}$ and ${\du},$ while $ f_{{\cu}{\au}{\du}}$ is totally antisymmetric.
As a result, eq. (\ref{vector1}) leads to
\be
\d_{(1)} \cV_{m \au} =  4{\rm i}\,  g^2\,  (\pa^n \s)\,
{\rm tr}_{\rm Ad} \,
\Big(T_{\au}{} \, (\tilde{\D}^{-3}|)  \, \cF_{nm} \Big) ~.
\label{vector2}
\ee

Since the expressions (\ref{scalar2}) and (\ref{vector2})
are already of first order in derivatives, we can approximate
the operator $\tilde{\D}$ defined in (\ref{tilde}) by
\be
\tilde{\D}_{ {\au} {\bu} } \approx -\pa^m \pa_m \, \d_{ {\au} {\bu} }
+ (\cM^2_{\rm v})_{ {\au} {\bu} }~, \qquad
 (\cM^2_{\rm v})_{ {\au} {\bu} } \equiv
 {\bar \cY} \{ T_{\au}, T_{\bu} \} \cY~,
\label{Mass-v}
\ee
with $\cY$ constant. Then, a direct evaluation of
(\ref{scalar2}) and (\ref{vector2}) gives
\bea
\d_{(1)} \cW_{\au} &=& -\frac{g^2}{8\p^2}\, (\pa^n \s)\,
{\rm tr}_{\rm Ad} \, ( T_{\au}{} \, \cM_{\rm v}^{-2} \cD_n \cW )~,
\label{scalar3} \\
\d_{(1)} \cV_{m \au} &=&
-\frac{g^2}{8\p^2}\, (\pa^n \s)\,
{\rm tr}_{\rm Ad} \,( T_{\au}{} \, \cM_{\rm v}^{-2} \cF_{nm} )~.
\label{vector3}
\eea
These relations determine the quantum modification to
the conformal transformations of the bosonic  fields
of the $\cN=2$ vector multiplet.

It is worth noting that the  mass matrix $\cM^2_{\rm v}$
may possess some zero eigenvalues, and therefore
$\cM_{\rm v}^{-2}$ is not defined by itself.
But in (\ref{vector3}), for example,
$\cM_{\rm v}^{-2}$ occurs in the combination
$\cM_{\rm v}^{-2} \cF_{nm}$
and the multiplier $\cF$
projects out all zero eigenvalues of $\cM_{\rm v}^{2}$.

${}$Finally, let us analyse the hypermultiplet conformal deformation.
We will use underlined Greek letters to denote
hypermultiplet indices, i.e. $Q_i = (Q_{i \underline{\a} }) $
and ${\bar Q}^i = ({\bar Q}^{i \underline{\a} }) $.
In accordance with (\ref{scalar1}), we have
\be
\d_{(1)} \cQ_{i \underline{\a} }
= 2{\rm i} \, (\pa^m \s)\,
(\tilde{\D}^{-1} )_{ {\bu} {\cu}} \,
<v_{ m {\cu} } (x) \, q_{i \underline{\b} } (x')>|_{x' = x} \,
(T_{\bu})_{ \underline{\a} }{}^{ \underline{\b} }~.
\label{scalar4}
\ee
To first order in the derivative expansion,
we can approximate $<v_m \, q_i>$ as
\bea
<v_{ m {\cu} } \, q_{i \underline{\b} } >
=2{\rm i}\, (\tilde{\D}^{-1})_{ {\cu} {\du} }
\Big\{ (T_{\du} \cD_m \cQ_j)_{ \underline{\gamma} }
&<{\bar q}^{j \underline{\gamma} } \, q_{i \underline{\b} }>_0 &
\non  \\
- (\cD_m {\bar \cQ}^j T_{\du})^{\underline{\gamma}}
&<q_{j \underline{\gamma} }  q_{i \underline{\b} }>_0 & \Big\}~.
\label{qvpropagator}
\eea
Here $<\ldots>_0$  denotes the propagator corresponding
to the free quadratic action
\bea
S_0 = \frac{1}{g^2} \int {\rm d}^4x \,
\Big\{  {\bar q}^i \pa^m \pa_m q_i
&-& {\bar q}^i \{ {\bar \cW}, \cW \} q_i \non \\
&+& ({\bar q}^i T_{\au} \cQ^j + {\bar \cQ}^j T_{\au} q^i ) \,
( {\bar q}_i T_{\au} \cQ_j + {\bar \cQ}_j T_{\au} q_i )
\Big\}~,
\eea
with $\cW$ and $\cQ_i$ constant.
Introducing
\bea
q= \left(
\begin{array}{c}
q_{j \underline{\b} } \\
{\bar q}^{j \underline{\b} }
\end{array}
\right) ~, \qquad
q^\dagger = ({\bar q}^{i \underline{\a} }, ~
q_{i \underline{\a} } )~,
\label{condensed}
\eea
the action can be rewritten in a more compact form
\be
S_0 = \frac{1}{2g^2} \int {\rm d}^4x \,
\Big\{ q^\dagger \pa^m \pa_m q
- q^\dagger \cM_{\rm h}^2 q \Big\} ~,
\label{hypermassmat}
\ee
where the hypermultiplet mass matrix is
\bea
\cM_{\rm h}^2 &=&
\left(
\begin{array}{ccc}
 \delta_i{}^j \,
\S_{\underline{\a}}{}^{\underline{\b}}
&  \vline &
- 2\epsilon_{ij} \,
(T_{\underline{a}} \cQ^k)_{\underline{\a}} \,
(T_{\underline{a}} \cQ_k)_{\underline{\b }}
\\
\hline
- 2\epsilon^{ij}
\, ({\bar \cQ}_k
T_{\underline{a}})^{\underline{\a}} \, ({\bar \cQ}^k
T_{\underline{a}})^{\underline{\b}}
& \vline &
\delta^i{}_j \,
\S_{\underline{\b}}{}^{\underline{\a}}
\end{array}
\right)\,,
\label{hyper-mass} \\
\S_{\underline{\a}}{}^{\underline{\b}} &=&
\{ {\bar \cW}, \cW \}_{\underline{\a}}{}^{\underline{\b}}
+ 2(T_{\underline{a}} \cQ_k)_{\underline{\a}} \, ({\bar \cQ}^k
T_{\underline{a}})^{\underline{\b}} ~. \non
\eea
Eq. (\ref{scalar4})  becomes
\bea
\d_{(1)} \cQ_{i \underline{\a} }
= -4  (\pa^m \s)\,
(T_{\bu})_{ \underline{\a} }{}^{ \underline{\b} }\,
(\tilde{\D}^{-2} )_{ {\bu} {\du}} \,
\Big\{ ~(T_{\du} \cD_m \cQ_j)_{ \underline{\gamma} } \,
<{\bar q}^{j \underline{\gamma} } \, q_{i \underline{\b} }>_0 &&\non \\
 \qquad - (\cD_m {\bar \cQ}^j T_{\du})^{\underline{\gamma}}
<q_{j \underline{\gamma} } \, q_{i \underline{\b} }>_0 &\Big\}&~.
\label{hyp-var-1}
\eea
Due to the off-diagonal terms in the
hypermultiplet mass matrix, the propagator\break
$<\!q_{j \underline{\gamma} } \, q_{i \underline{\b}}\!>_0$
does not vanish in general. For the background field configurations
we will consider below, it does, however, vanish.

Unlike eqs. (\ref{scalar3}) and (\ref{vector3}),
the variation $\d_{(1)} \cQ$ involves two different propagators
with different mass matrices, $\cM_{\rm v}^2$ and $\cM_{\rm h}^2$.
The transformation (\ref {hyp-var-1}) for the hypermultiplets
will have a form  similar to (\ref{scalar3}) only if a
special relationship exists
between the mass matrices $\cM_{\rm v}^2$ and $\cM_{\rm h}^2$.
The point is that $\d_{(1)} \cQ$ proves to be a linear combination
of terms proportional to the following (Euclidean) momentum integral
\bea
\int \frac{ {\rm d}^4 k }{ (2\p)^4 } \,
\frac{1}{(k^2 +M_1{}^2)^2} \, \frac{1}{(k^2 +M_2{}^2)}
&=&  \frac{M_2{}^2}{32\p^2 (M_1{}^2 - M_2{}^2)^2} \,
\ln \Big(\frac{M_2{}^2}{M_1{}^2} \Big) \non \\
&+& \frac{1}{32\p^2 (M_1{}^2 - M_2{}^2)}~,
\label{massintegral}
\eea
with $M_1{}^2$ and $M_2{}^2$ being  some eigenvalues of
$\cM_{\rm v}^2$ and $\cM_{\rm h}^2$ respectively.
${}$Only for $M_1 =M_2 =M$ is the mass dependence of the form
$1/(32 \p^2 M^2)$, as in (\ref{scalar3}).
This therefore raises the question: under what circumstances are
there coinciding mass eigenvalues?

Let the background vector multiplet fields $(\cV_m,  \cW)$
be of the form (\ref{separ-of-variables}), with $H$ a given generator
in the Cartan subalgebra.
It is useful to examine the implications of
the $U(1)$ gauge symmetry (\ref{back-trans}) generated by $H$.
We know that (i) the background scalar fields $\cW$ and
$\cQ_i$ are invariant under the $H$-gauge transformations;
(ii) the quadratic action (\ref{quadrat-action})
is invariant under the $H$-gauge transformations.
These observations imply
\be
[H, \cM^2_{\rm v} ] =0~, \qquad [H, \cM^2_{\rm h} ] =0~,
\label{charge-mass}
\ee
and therefore the charge operator $H$ and the mass matrices
can be simultaneously diagonalized.
One further observes that
the derivative interaction terms in  (\ref{quadrat-action}),
\bea
\frac{\rm i}{2}\, L_{\rm int}^{(1)} &=&
{\bar w} (\cD_m \cW) v^m -v^m (\cD_m {\bar \cW} ) w
-\hf v^m \cF_m{}^n  v_n ~,
\label{vert-1}\\
\frac{\rm i}{2} L_{\rm int}^{(2)}\,  &=&
 (\cD_m {\bar \cQ}^i) v^m q_i -{\bar q}^i v^m \cD_m \cQ_i ~,
\label{vert-2}
\eea
are invariant under the $H$-gauge transformations.
The $H$-invariance of
$L_{\rm int}^{(1)}$ implies that each term on
the right-hand side of $L_{\rm int}^{(1)}$
involves two quantum fields of opposite $H$-charge;
these fields have the same mass due to the first identity
in (\ref {charge-mass}). Similarly, the $H$-invariance of
$L_{\rm int}^{(2)}$ implies that each term on
the right-hand side of $L_{\rm int}^{(2)}$
involves two quantum fields of opposite $H$-charge.
In order for these fields to have the same mass,
it is sufficient to require
(using the condensed notation (\ref{condensed}))
\be
(\cD_m \cQ^\dagger) \, v^m \, \cM^2_{\rm h} \,q
=  (\cD_m \cQ^\dagger) \, (\cM^2_{\rm v} v^m ) \, q  ~,
\ee
as a consequence of (\ref {charge-mass}).
The latter requirement is satisfied provided $\cQ_i$
is of  the form (\ref{separ-of-variables}).
In this case, the off-diagonal terms in the hypermultiplet mass
matrix (\ref {hyper-mass})  also vanish.

Thus, when the conditions (\ref{separ-of-variables}) are imposed,
 considerable simplification of the expression
(\ref{hyp-var-1}) occurs. The vector mass matrix $\cM_{\rm v}^2$ can
be diagonalized by an appropriate choice of basis for the Lie algebra
of the gauge group. Each non-zero mass eigenvalue corresponds to a
generator $T_{\hat{a}}$ which is broken, in the sense that $[T_{\hat{a}},
\cW] \neq
0$ and/or $T_{\hat{a}} \cQ_i \neq 0.$ The deformation
(\ref{hyp-var-1}) can be put in the form
\be
\d_{(1)} \cQ_{i } = -\frac{g^2}{8 \pi^2} \, (\pa^m \s)\,
\sum_{I}
\frac{1}{M_{I}{}^2} \, {\sum_{\hat{a}}}^{\prime}\,  T_{\hat{a}} T_{\hat{a}}
\cD_m
\cQ_i~.
\label{simpleh}
\ee
Here,  $I$ labels the different nonzero eigenvalues $M_{I}{}^2$  of the
vector mass matrix
$\cM_{\rm v}^2$, and, for a given $I,$ the
sum $\sum^\prime_{\hat{a}}$ is
over all broken generators $T_{\hat{a}}$ corresponding to the mass
eigenvalue $M_I{}^2.$  To derive this result, one
makes use of the fact that when the conditions
(\ref{separ-of-variables}) are imposed, the propagator
$<q_{j \underline{\gamma} } \, q_{i \underline{\b} }>_0$
in (\ref{qvpropagator}) vanishes, as there are no off-diagonal blocks in the
hypermultiplet mass matrix. Further, as proven above,  the
interaction $(T_{{\du}} \cD_m \cQ_j)_{\underline{\g}}$
in the first term on the right-hand side of (\ref{qvpropagator})
only mixes hypermultiplets and vectors of the same mass. Thus the
propagator (\ref{qvpropagator}) decomposes into a sum of terms, one
for each non-zero mass eigenvalue of the vector mass matrix. For a
given mass eigenvalue $M_{I}{}^2,$ the operator
$ (\tilde{\Delta}^{-1}
)_{\hat{c} \hat{d} } $
takes the form
$(-\partial^2 +
M_{I}{}^2)^{-1} \,
\delta_{\hat{c} \hat{d} },
$ while
\be
<\bar{q}^{j \underline{\gamma} } \, q_{i \underline{\b} }>_0 = -
{\rm i} \, g^2 \,
(-\partial^2 +
M_{I}{}^2)^{-1} \, \delta^j{}_i \,
\delta^{\underline{\gamma}}{}_{ \underline{\b}}~.
\ee
Substituting the resulting
expression (\ref{qvpropagator}) into
(\ref{scalar4}),
the deformed hypermultiplet transformation is
\be
\d_{(1)} \cQ_{i } =  4 \, {\rm i} \,  g^2 \, \, (\pa^m \s)\,
\sum_{I}
(-\partial^2 + M_I{}^2)^{-3}| \, {\sum_{\hat{a}}}^{\prime}\,  T_{\hat{a}}
T_{\hat{a}}
\cD_m
\cQ_i~.
\ee
In momentum space, this yields integrals of the form
(\ref{massintegral}) with $M_1{}^2 = M_2{}^2 = M_I{}^2.$

An analogous result applies
for the transformations (\ref{scalar3}),
which can be cast in the form
\be
\d_{(1)} \cW = -\frac{g^2}{8\p^2}\, (\pa^m \s)\,
\sum_{I}
\frac{1}{M_{I}{}^2} \, {\sum_{\hat{a}}}^{\prime} \, [T_{\hat{a}}, \,
[T_{\hat{a}}
,\cD_m \cW ]] ~,
\label{simplea}
\ee
and similarly for the vector transformation (\ref{vector3}).
Of course, the covariant derivatives in (\ref{simpleh})
and (\ref{simplea}) coincide with partial derivatives
under the (gauge) choice (\ref{separ-of-variables}).

\sect{Conformal deformation on the Coulomb branch}

Relations (\ref{scalar3}) and (\ref{vector3})
allow us to evaluate the one-loop deformation of
conformal symmetry on the Coulomb branch, where we restrict the analysis in
this section to the case where the scalars in the
hypermultiplets are zero\footnote{This is the generic case,
but e.g. the model discussed in sect.~6 also allows for non-vanishing
$\cQ_i$ without reducing the rank of the gauge group.}, i.e. we
consider solutions to the equations
\be
[{\bar \cW}, \cW] = 0~, \qquad  \cQ_i = 0~.
\label{flat-coul}
\ee
In accordance with (\ref{Mass-v}), the mass matrix
$\cM^2_{\rm v}$ now becomes
\be
\hf \, \cM^2_{\rm v}\, \z = [{\bar \cW}, [\cW, \z]]
= [\cW , [{\bar \cW}, \z]]~.
\ee
${}$For definiteness, the variations
(\ref{scalar3}) and (\ref{vector3})  will be evaluated
for a special
superconformal theory -- $\cN=2$ super Yang-Mills
with gauge group $SU(N)$ and $2N$ hypermultiplets
in the fundamental \cite{HSW}. This example
captures the general features which we wish to highlight;
in particular the analysis of this section is directly applicable
to the models considered in sect.~6 and 7 after setting the
hypermultiplets to zero.
It is worth noting that the finiteness condition
(\ref{finiteness}) for the $SU(N)$ model follows from the following
identity for the group $SU(N)$:
${\rm tr}_{\rm Ad}  = 2N\, {\rm tr}_{\rm F}$.

${}$First we briefly recall some well known facts concerning
the explicit structure of the Coulomb branch of
the theory under consideration (see, for instance,
\cite{APS}). Up to a gauge transformation, the general solution
to the first equation in (\ref{flat-coul}) is
\be
\cW = {\rm diag} (W_1, \ldots , W_N )~, \qquad
\sum_{\a} W_\a  =0~.
\ee
Generically, when all eigenvalues of $\cW$
are different, the gauge symmetry $SU(N)$ is broken to
$U(1)^{N-1}$; then, $\cQ=0$ is the only solution to the
second and third equations in (\ref{flat}).
If $k$ eigenvalues of $\cW$ coincide, the gauge group $SU(N)$
is broken to $SU(k) \times U(1)^{N-k}$; the second and third equations
in (\ref{flat}) have non-trivial solutions when some eigenvalues of
$\cW$ vanish.

Here we will be interested in a maximally symmetric
field configuration which contains a single Abelian
$\cN=2$ vector multiplet,
\be
\cW = \bW \, H~, \quad
\cV_m = \bV_m \, H~, \quad
H =  \frac{1}{\sqrt{N(N-1)}} \,
{\rm diag}\, ( N-1, -1,  \cdots, -1) ~,
\label{maxsym}
\ee
and leaves the subgroup $SU(N-1) \times U(1)\subset SU(N)$ unbroken.
With this choice for $\cW$ the only solution to the
second equation in (\ref{flat}) is $\cQ=0$.

${}$For the $\cN=2$ vector multiplet (\ref{maxsym}),
the explicit structure of
the variations  (\ref{scalar3}) and (\ref{vector3})
can be read off from similar results for the
$\cN=4$ super Yang-Mills theory \cite{KM1},
see also sect. 5. One obtains
\bea
 \d_{(1)} \bW =
-\frac{g^2 (N-1)}{8 \pi^2} \,
\frac{ (\partial^n \sigma) \,
\partial_n \bW }
{ {\bar \bW} \bW } ~,
\qquad
\d_{(1)} \bV_m =
-\frac{g^2 (N-1)}{8 \pi^2} \,
\frac{ (\partial^n \sigma) \,
\bF_{nm} }
{ {\bar \bW} \bW } ~  .
 \label{CB}
\eea
This result is universal: modulo an overall common factor,
which is the number of broken generators,
the same deformation of conformal symmetry occurs
for any $\cN=2$ superconformal theory in the Coulomb phase
described by a single $U(1)$ vector multiplet.

The deformed conformal transformations  (\ref{multi-loop-deform}),
with the quantum corrections from all loops  and
all orders in the derivative expansion taken into account,
should realize the conformal algebra,
as can be justified by a direct generalization of
the considerations given in \cite{Pal,FP2}.
On the Coulomb branch, the one-loop quantum deformation
to first order in derivatives,
\bea
\d \bV_m &=& \d_{\rm c} \bV_m - R^4\,
\frac{ (\partial^n \sigma) \, \bF_{nm} }
{ 4{\bar \bW} \bW } ~, \label{CB-v1}\\
\d \bW &=& \d_{\rm c} \bW - R^4  \,
\frac{ (\partial^n \sigma) \,\partial_n \bW }
{ 4{\bar \bW} \bW } ~,  \label{CB-s1}
\eea
with $R^4 = g^2(N-1)/ 2\p^2$,
realizes the conformal algebra (up to a pure gauge transformation)
without the need to
take any higher loop corrections into account.
Using the non-renormalization theorem of
Dine and Seiberg \cite{DS}, we will show
that the above conformal deformation is one-loop exact.

It should first be pointed out that the transformations
(\ref{CB-s1}) and (\ref{CB-v1}) coincide with the
$AdS_5$ isometries of a (static gauge) D3 brane
embedded in $AdS_5 \times S^1$:
\bea
&&S = - {1 \over g^2} \int {\rm d}^4x
\left( \sqrt{ - {\rm det} \Big( \frac{\bX^2}{R^2} \, \eta_{m
n} +  \frac{R^2}{\bX^2}\, \partial_{m}\bX_\m \partial_{n} \bX_\m
+ \bF_{mn} \Big)} - \frac{\bX^4}{R^4}~
\right) ~,
\label{d3brane} \\
&&
\qquad \qquad \qquad \qquad
\bX^2 = \bX_\m \bX_\m ~,
\non
\eea
where $\bX^2  =2\,{\bar \bW} \bW$ and $\bX_\m = (\bX_1, \bX_2)$ are defined
by $\sqrt{2} \, \bW = \bX_1 +{\rm i} \, \bX_2$.
Of course, the conformal invariance (\ref{CB-s1}) and (\ref{CB-v1})
does not uniquely fix the D3 brane action. But in conjunction
with $\cN=2$ supersymmetric non-renormalization theorems,
it severely restricts the functional form of invariant actions.

Let us demonstrate, by analogy with Maldacena's analysis
\cite{Maldacena},  that the conformal symmetry  (\ref{CB-s1})
and simple $\cN=2$ non-renormalization theorems allow one
to restore part of
the low energy effective action of the form
\be
\G[\bW, {\bar \bW}] = \int {\rm d}^4 x \,L_{\rm eff}
(\bW, {\bar \bW},\pa \bW, \pa {\bar \bW} )~.
\ee
It is useful to introduce radial $\bX$ and angular $\vf$ variables,
$\sqrt{2} \, \bW = \bX \exp ({\rm i} \,\vf)$. It follows from
(\ref{CB-s1}) and (\ref{CB-v1}) that $\ln \bX$ can be interpreted as
a Goldstone field for partial symmetry breaking of the conformal group
$SU(2,2)$ to $SO(4,1)$. Similarly, $\vf$ can be treated as
a Goldstone field corresponding to spontaneous breakdown of the
$U(1)_\cR$ factor in the $\cN=2$ $\cR$-symmetry group
$U(1)_\cR \times SU(2)_\cR$, of which the $SU(2)_\cR$ factor  leaves
$W$ invariant.
As follows, for instance, from the techniques of nonlinear realizations,
the general form for $L_{\rm eff}$ is
\be
L_{\rm eff} = \g \, \bX^4
+\sqrt{-G} \, \sum_{k=0}^{\infty} c_{2k} \,
(G^{mn}\, \pa_m \vf \pa_n \vf)^k ~, \qquad
G = \det (G_{mn})~,
\label{L-eff}
\ee
with $\g$ and $c_k$ constant parameters,
and $G_{mn}$ the induced  metric on the D3-brane,
\be
G_{mn} = \frac{\bX^2}{R^2} \, \eta_{m
n} +  \frac{R^2}{\bX^2}\, \partial_{m}\bX_\m \partial_{n} \bX_\m
= \frac{\bX^2}{R^2} \, \eta_{m
n} +  \frac{R^2}{\bX^2}\, \partial_{m}\bX \partial_{n} \bX
+R^2 \, \partial_{m} \vf \partial_{n} \vf ~.
\ee
Of course, the numerical coefficient of the angular term,
$\pa_m \vf \pa_n \vf$, in $G_{mn}$ can be changed at the
expense of modifying the infinite series in (\ref{L-eff}).
The $\bF$-independent part of the D3 brane Lagrangian in (\ref{d3brane})
is of the form (\ref{L-eff}) with $c_0= - R^4 \,\g = - 1/ g^2$
and $c_{2k} = 0$ for $k \geq 1$.

The theory we are considering is $\cN=2$ superconformal,
and thus neither the kinetic term nor the scalar potential
receive quantum corrections.
These non-renormalization properties and the fact that
the field configuration (\ref{maxsym})
corresponds to a flat direction imply that the part
of $L_{\rm eff}$  with at most two derivatives must
coincide with the tree-level Lagrangian, and hence
\be
L_{\rm eff} = -{1 \over g^2} \, \pa^m {\bar \bW} \pa_m \bW ~+~
O(\pa^4)~,
\ee
implying that $c_0= - R^4 \,\g = - 1/ g^2$
and $c_2 = 0$. We see that the radial part of $L_{\rm eff}$
is completely fixed by the conformal symmetry  (\ref{CB-s1})
and $\cN=2$ non-renormalization theorems \cite{Maldacena}.
In addition, the Dine-Seiberg theorem \cite{DS}
turns out to imply $c_4=0$. Let $W (x, \q )$ be the $\cN=2$
chiral superfield strength with $\bW(x)$ being its leading
($\q$-independent) component. The non-renormalization theorem
of Dine and Seiberg states that, in the Coulomb branch
of $\cN=2$ superconformal theories,  the following
manifestly $\cN=2$ supersymmetric quantum correction
to the low energy effective action
\be
c  \int {\rm d}^4 x \,{\rm d}^4 {\bar \q} {\rm d}^4 \q \,
\ln {\bar W}  \ln W
\label{F-4}
\ee
is one-loop exact. In components, this functional
contains two special bosonic structures with four derivatives
(with $\bF_{\a \b}$ and ${\bar \bF}_{\ad \bd}$ the helicity $+1$
and $-1$ components of the field strength $\bF_{mn}$):
\be
\frac{\bF^{\a \b} \bF_{\a \b}\,
{\bar \bF}^{\ad \bd} {\bar \bF}_{\ad \bd} }
{( {\bar \bW} \bW)^2}~,  \qquad \quad
\frac{(\pa {\bar \bW})^2 \, (\pa \bW)^2}
{( {\bar \bW} \bW)^2}~,
\ee
which precisely match the four-derivative structures
appearing in the expansion of the Born-Infeld action (\ref{d3brane}),
including the relative coefficient \cite{GKPR}.
The explicit value of the overall
coefficient $c$ in (\ref{F-4}) for the
theory under consideration was computed in
\cite{GKPR,BBK,LvU} to be $c= (N-1)/ (4\p)^2$.
This confirms that $c_4 =0$.
The coefficients $c_6, c_8,\ldots,$ in (\ref{L-eff})
may have non-vanishing values, in particular $c_6 \neq 0$.
The point is that the $\bF^6$ quantum correction in the low energy
effective action 
of the $SU(N)$ gauge theory with $2N$ fundamental hypermultiplets
is known to be different from the one coming
from the Born-Infeld action (\ref{d3brane}) \cite{BKT},
and the leading part of the $c_6$ term  in (\ref{L-eff}) should
occur in the same superfield functional which contains those
$\bF^6$ terms which are not present in the Born-Infeld action.

In summary, the terms in $L_{\rm eff}$ containing two and four
derivatives of the fields coincide with the corresponding terms in
the D3-brane action (\ref{d3brane}), and the four-derivative term
in $L_{\rm eff}$ is one-loop exact. This information suffices
to argue that the parameter $R^4$ in
(\ref{CB-s1}) and (\ref{CB-v1})  cannot receive two- and higher-loop
quantum corrections.
Indeed, the transformation (\ref{CB-s1}) mixes the
two-derivative and four-derivative terms in $L_{\rm eff}$,
of which the former is tree-level exact and the latter
is one-loop  exact.

\sect{\mbox{$\cN=4$ SYM}}

In the case of $\cN=4$ SYM,
which is just a special $\cN=2$ SYM theory,
the hypermultiplet sector is
composed of a single hypermultiplet in the adjoint
representation of the gauge group, and hence
$Q_{i \underline{\a} } =  Q_{i \underline{a} }$
and ${\bar Q}^{i \underline{\a} }={\bar Q}^i_{ \underline{a} }$.
With the use of eq. (\ref{flat}), one can now show that
the off-diagonal elements of the hypermultiplet
mass matrix $\cM_{\rm h}^2$ vanish,
\be
(T_{\underline{c}} \cQ^k)_{\underline{a}} \,
(T_{\underline{c}} \cQ_k)_{\underline{b }} =0~,
\ee
while the block diagonal pieces of $\cM_{\rm h}^2$
coincide with the vector multiplet
mass matrix, $\cM^2_{\rm v}$,
\be
\{ {\bar \cW}, \cW \}_{\underline{a} \underline{b}}
+ 2(T_{\underline{c}} \cQ_k)_{\underline{a}} \, ({\bar \cQ}^k
T_{\underline{c}})_{\underline{b}}
= {\bar \cW} \{ T_{\au}, T_{\bu} \} \cW
+{\bar \cQ}^i \{ T_{\au}, T_{\bu} \} \cQ_i
= (\cM^2_{\rm v})_{ {\au} {\bu} }~.
\ee
As a consequence, eq. (\ref{hyp-var-1}) becomes
\be
\d_{(1)} \cQ_{i \underline{a} } =
 4{\rm i}\,  g^2\,  (\pa^m \s)\,
{\rm tr}_{\rm Ad} \,
\Big(T_{\au}{} \, (\tilde{\D}^{-3}|)  \, \cD_m \cQ_i \Big) ~,
\label{hyp-var-2}
\ee
which is of the same form as (\ref{scalar2}) and (\ref{vector2}).

Evaluating (\ref{hyp-var-2}), the result can be combined
with eqs. (\ref{scalar3})  and (\ref{vector3})
to give the final expression for the one-loop
deformation of conformal symmetry:
\bea
\d_{(1)} \cV_{m \au} &=&
-\frac{g^2}{8\p^2}\, (\pa^n \s)\,
{\rm tr}_{\rm Ad} \,( T_{\au}{} \, \cM_{\rm v}^{-2} \cF_{nm} )~,
\non \\
\d_{(1)} \cW_{\au} &=& -\frac{g^2}{8\p^2}\, (\pa^n \s)\,
{\rm tr}_{\rm Ad} \, ( T_{\au}{} \, \cM_{\rm v}^{-2} \cD_n \cW )~,
\label{n4final}
\\
\d_{(1)} \cQ_{i {\au}} &=&
-\frac{g^2}{8\p^2}\, (\pa^n \s)\,
{\rm tr}_{\rm Ad} \,( T_{\au}{} \, \cM_{\rm v}^{-2} \cD_n \cQ_i )~.
\non
\eea
By construction, the mass matrix $\cM^2_{\rm v}$
is invariant under the $\cR$-symmetry group $SO(6)_\cR$ which rotates
the six scalars $\cW$, $\bar \cW$, $\cQ_i$ and ${\bar Q}^i$.
It is clear that the one-loop deformation (\ref{n4final})
respects the $SO(6)_\cR$ symmetry.

We want to evaluate (\ref{n4final}) for a general
semi-simple gauge group $G$.
For this purpose it is convenient to use a complex basis
for the charged gauge bosons and their scalar partners. We choose them in
one-to-one correspondence with the roots of the Lie-algebra.
That is $V=\sum_{\au}V^{\au}T_{\au}=
\sum_\a V^\a E_\a+\sum_i V^i H_i$
and likewise for the adjoint scalars. Here
$E_\a$ is the generator corresponding to the root $\a$
normalized as $\tr(E_\a E_\b^\dagger)=\tr(E_\a E_{-\b})=\delta_{\a,\b}$,
and $H_i$ are the rank($G$) Cartan subalgebra generators. They satisfy the
commutation relations $[H_i,E_\a]=\a(H_i) E_\a$.
With all background fields aligned along an arbitrary element $H$ of
the Cartan subalgebra (cf. (\ref{separ-of-variables}) with
$\U=H$),
the vector-multiplet mass matrix becomes
\be
({\cal M}_{\rm v}^2)_{\a\b}={\bf X}^2 \,\tr([H,E_\a^\dagger][E_\b,H])
={\bf X}^2 \,(\a(H))^2\delta_{\a,\b}
\label{vmassm}
\ee
with ${\bf X}^2=2(\bar{\bf W}{\bf W}+\bar{\bf Q}^i{\bf Q}_i)$.
All components of ${\cal M}_{\rm v}^2$ along the Cartan subalgebra
directions vanish:
the neutral gauge bosons stay massless (adjoint breaking does not reduce the
rank of the gauge group) and they do not mix with the charged vector bosons
at the quadratic level.

It is now straightforward to compute, say, $\delta_{(1)}{\bf W}$, for
which we get from (\ref{n4final})
$\delta_{(1)}{\bf W}\propto{\rm tr}_{\rm Ad}(H^2 {\cal M}_{\rm v}^{-2})$.
${}$From the commutation relation $[H,E_\a]=\a(H)E_\a$ we read off
$H$ in the adjoint representation. Then the factors $\a(H)$ cancel
and the trace produces a numerical coefficient equal to the number of
roots with $\a(H)\neq0$, i.e. the number of broken generators or,
in other words, the number of massive gauge bosons.

To be specific,
we consider the classical groups with the background chosen such
that one obtains the breaking patterns $SU(N)\to SU(N-1)\times U(1),\,
Sp(2N)\to Sp(2N-2)\times U(1),\,SO(N)\to SO(N-2)\times U(1)$.
These regular subgroups are obtained by removing an extremal
node from the Dynkin diagram of the gauge group.
In the brane realization of these theories this corresponds to moving
one of the D3 branes away from the orientifold O3 plane.
This implies a particular Cartan subalgebra generator $H$
(for $SU(N)$ it has been explicitly given in (\ref{maxsym}))
for which $\a(H)=\a\cdot\mu$, where $\mu$ is the fundamental weight
corresponding to the simple root associated with the removed node of the Dynkin
diagram.
We then find,
\be
\delta_{(1)}{\bf W}= -{\l g^2\over 8\pi^2 \bX^2}
(\partial^n\s)\partial_n{\bW}\,,\quad
\hbox{with}\quad
\l=
\cases{
2(N-1)&{\rm for} $SU(N)$,\cr
4N-2&{\rm for} $Sp(2N)~{\rm and}~SO(2N+1)$,\cr
4N-1&{\rm for} $SO(2N)$,\cr}
\label{d1W}
\ee
and likewise for $\delta_{(1)}{\bf Q}_i$ and
$\delta_{(1)}{\bf V}_m$.
As mentioned above, $\l$ is simply the number of broken generators.
In the  $SU(N)$ case, the deformation $\delta_{(1)}{\bf W}$
was derived in \cite{JKY,KM1}.

What remains to be shown is that the other components of the
background fields do not receive any  one-loop deformation,
i.e. that the deformation vanishes for $T_{\au}$ such that
${\rm tr}(T_{\au} H)=0$. Indeed, for any direction $\au$ the
deformation is proportional to $\sum_{\a}{1\over\a(H)}(T_{\au})_{\a\a}$,
where the sum is restricted to those roots for which $\a(H)\neq0$.
This vanishes if $T_{\au}$ is one of the $E_\a$ (since
$[E_\a,E_\b]$ is never proportional to $E_\a$). If $T_{\au}=H'$
with ${\rm tr}(H H')=0$
the deformation is proportional to
$\sum_\a{\a(H')\over\a(H)}$.
Using the roots and fundamental weights for the classical Lie algebras
(see e.g. the Appendix of \cite{Bourbaki})
this can be explicitly shown to vanish.

The results of this section easily generalize to ${\cal N}=2$ theories
on the pure Coulomb branch by simply
setting ${\bX}^2=2 \bar{\bW}\bW$.

It is worth discussing the results obtained in this section.
To first order in the derivative expansion,
the one-loop deformed conformal transformations are:
\bea
\d  \bV_m &=& \d_{\rm c} \bV_m - R^4 \,
\frac{ (\partial^n \sigma) \, \bF_{nm} }
{4({\bar \bW} \bW +{\bar \bQ}^j \bQ_j)} ~, \non \\
\d \bW &=& \d_{\rm c} \bW - R^4  \,
\frac{ (\partial^n \sigma) \,\partial_n \bW }
{ 4({\bar \bW} \bW +{\bar \bQ}^j \bQ_j)} ~;  \label{n=4dct}\\
\d \bQ_i &=& \d_{\rm c} \bQ_i - R^4  \,
\frac{ (\partial^n \sigma) \,\partial_n \bQ_i }
{4( {\bar \bW} \bW +{\bar \bQ}^j \bQ_j)} ~, \non
\eea
with $R^4 = \l g^2/ 4 \p^2$ and $\l$ defined as above
for the classical groups. These transformations
leave invariant the D3 brane action (\ref{d3brane}),
which now involves six scalars $\bX_\m$, where $\m =1,\ldots,6$,
defined by $\sqrt{2} \, \bW = \bX_1 +{\rm i} \, \bX_2$ and
$\sqrt{2} \, \bQ_i = \bX_{i+1} +{\rm i} \, \bX_{i+2}$.
The transformations (\ref{n=4dct}) realize the conformal
algebra (up to a pure gauge transformation)
without the need to take into account any higher
loop quantum corrections.
The consideration of sect. 4 implies that the  parameter
$R^4$ in (\ref{n=4dct}) is one-loop exact.
Unlike the situation with generic $\cN=2$ superconformal theories
on the Coulomb branch, which we discussed in sect. 4,
the low energy effective action in the $\cN=4$ SYM theory
is expected to be of the Born-Infeld form (\ref{d3brane}),
at least in the large $N$ limit (see \cite{CT,Tseytlin}
and references therein). For this to hold,\footnote{The claim
\cite{KM1,KM2} that the deformed conformal symmetry
(\ref{n=4dct}), $SO(6)_\cR$ invariance
and the {\it known} $\cN=4$ non-renormalization
theorems (which we discussed in sect. 4 of the present paper)
uniquely fix the scalar part,
$L_{\rm eff} (\bX_\m, \pa_m \bX_\n)$,
of the low energy effective Lagrangian in $\cN = 4$ SYM, is incorrect.}
there should exist a host of (yet unknown) non-renormalization theorems
in $\cN=4$ SYM (see, e.g., the discussion in \cite{BKT}).

\sect{\mbox{$USp(2N)$} SYM with fundamental and traceless \\
antisymmetric hypermultiplets}

Here we consider the $\cN=2$ superconformal Yang-Mills theory
introduced in \cite{ASTY,DLS}, which is known to have
a supergravity dual on $AdS_5 \times S^5 / \bZ^2$
\cite{Fayyazuddin}.
It will be shown that the structure of the quantum conformal
deformation differs significantly from the $\cN=4$ SYM case.

The gauge group is $USp(2N) = Sp(2N, {\bC})
\bigcap U(2N)$, and the theory contains hypermultiplets
in two representations of the gauge group:
four hypermultiplets\footnote{The $SU(2)_\cR$
indices of the hypermultiplets are suppressed.}
$Q_{\rm F}$ in the fundamental and
one hypermultiplet $Q_{\rm A}$
in the antisymmetric traceless representation of $USp(2N)$.
The Lie algebra $usp(2N)$ is spanned by $2N \times 2N$
matrices $({\rm i}\, \z) $ satisfying the  constraints
\bea
\z^{\rm T}\, J + J \,\z = 0~, \qquad  \z^\dagger = \z~, \qquad
J =- J^{\rm T} = \left(
\begin{array}{cc}
0 & {\bf 1}_N \\
-{\bf 1}_N & 0
\end{array}
\right)~.
\eea
We  will use
Greek letters to denote the components of $USp(2N)$
spinors, $Q_{\rm F} = ( Q_{{\a}}) $,
and also make use of the symplectic metric
$J= (J^{{\a \b} })
= (J_{{\a \b} })$
for raising and lowering $USp(2N)$ spinor indices;
for example, $\J^{{\a}} \equiv J^{{\a \b} }
\J_{{\b}} $ and
$\J_{{\a}} \equiv - J_{{\a \b} }
\J^{{\b}}{}$. The antisymmetric traceless representation
of $USp(2N)$ is realized by second rank tensors of the form
\be
Q_{\rm A} = ( Q_{{\a}}{}^{{\b}} )~,
\qquad {\rm tr}\,  Q_{\rm A} =0~, \qquad
Q_{{\a \b} } = - Q_{{\b \a} }~.
\ee
The hypermultiplets $Q_{\rm F}$ and $Q_{\rm A}$ transform
under $USp(2N)$ as
\be
\d Q_{\rm F} ={\rm i} \,\z \,  Q_{\rm F}~,
\qquad
\d Q_{\rm A} ={\rm i} \,[ \z ,   Q_{\rm A}] ~.
\ee
It is worth noting that the symmetric representation of
$USp(2N)$ can be identified with the Lie algebra $sp(2N, {\bC})$,
i.e. $ \z = ( \z_{{\a}}{}^{{\b}} ) \in sp(2N, {\bC})$
iff $\z_{{\a \b} } = \z_{{\b \a} } $.
We also note that the finiteness condition (\ref{finiteness})
is met due to the following property of $USp(2N)$ representations:
\be
{\rm tr}_{\rm Ad} W^2 - 4\, {\rm tr}_{\rm F} W^2
-{\rm tr}_{\rm A} W^2 = 0~.
\ee

The background fields $\cW$, $\cQ_{\rm F}$ and $\cQ_{\rm A}$ are
chosen to solve  the equations (\ref{flat}) defining the classical moduli
space. Up to a gauge transformation, the general solution to the
first equation  in  (\ref{flat}) is
$\cW = {\rm diag} \,(W_1,  W_2, \ldots , W_N ,
- W_1, -W_2,  \ldots , -W_N )$,
with the $W$'s arbitrary complex numbers.
Further analysis is restricted to the choice
$W_1 \neq 0$ and
$W_2= \ldots = W_N=0;$ in this case, the
unbroken gauge subgroup, $USp(2N -2) \times U(1)$, is maximal.
We will also impose the requirement that
$\cQ_{\rm F}$ and $\cQ_{\rm A}$,
which must be solutions to the second and third equations in
(\ref{flat}),  be invariant  under the unbroken group
$USp(2N - 2) \times U(1)$. This leads to\footnote{The background
fields are chosen to yield correctly  normalized 
kinetic terms in the classical action, 
$-{1 \over g^{2}}\int {\rm d}^4x \, (\pa^m {\bar \bW} \pa_m \bW 
+ \pa^m {\bar \bQ}^i \pa_m \bQ_i)$.}
\bea
\cW &=& \frac{\bW}{\sqrt{2} }  ~
{\rm diag} \,(1,
\underbrace{ 0, \ldots, 0 }_{N-1}, -1,
\underbrace{ 0, \ldots, 0}_{N-1})~,  \\
\cQ_{\rm F} = 0~, \quad
\cQ_{\rm A} &=& \frac{\bQ}{\sqrt{2 N (N-1)} }  ~
{\rm diag} \,(N-1,
\underbrace{ -1, \ldots, -1 }_{N-1}, N-1,
\underbrace{ -1, \ldots, -1 }_{N-1})~.\non 
\eea
The nonvanishing background fields  constitute
the bosonic sector of an Abelian $\cN=2$ vector multiplet,
$(\bV_m, \bW, {\bar \bW})$, and a single neutral hypermultiplet,
$( \bQ_i , {\bar \bQ}^j)$.
Since the background is of the form
(\ref{separ-of-variables}), the formulas (\ref{simpleh}) and
(\ref{simplea}) can be applied to determine one-loop deformations of
the conformal
transformations once the mass matrices are known.

The mass matrix $\cM_{\rm v}^2$ for quantum
fields in the adjoint representation has $4(N-1)$ eigenvectors
with the eigenvalue $ {\bar \bW}\bW + \frac{N}{N-1}{\bar \bQ}^j \bQ_j$
and two eigenvectors with the eigenvalue $4{\bar \bW} \bW$.
The massive degrees of freedom in the quantum adjoint scalars  $w$
are parametrized in the form
\be
w = {1 \over \sqrt{2} }\,
\left(\begin{array}{cc}
Z & R \\
S & -Z^{\, \rm T}
\end{array} \right)
\ee
with
\be
Z = \left( \begin{array}{cl}
0 &~ \vec{z}_2^{\, \rm T}  \\
\vec{z}_1 & ~{\bf 0}_{N-1}
\end{array}\right)~, \quad
R = \left( \begin{array}{cl}
\sqrt{2} \,\a &~  \, \vec{r}^{\, \rm T}  \\
\vec{r} & ~{\bf 0}_{N-1}
\end{array}\right)~, \quad
S = \left( \begin{array}{cl}
\sqrt{2} \,\b & ~ \,\vec{s}^{\, \rm T}  \\
\vec{s} & ~{\bf 0}_{N-1}
\end{array}\right)~,
\ee
where $\vec{z}_1,$
$\vec{z}_2,$ $\vec{r}$ and $\vec{s}$ are ($N-1$)-vectors.
The fields which form the components of
$\vec{z}_1,$ $\vec{z}_2,$ $\vec{r}$ and $\vec{s}$
are eigenvectors of $\cM_{\rm v}^2$
with the eigenvalue $ {\bar \bW} \bW + \frac{N}{N-1}{\bar \bQ}^j \bQ_j,$
while the scalars $\a$ and $\b$  are eigenvectors of $\cM_{\rm v}^2$
with the eigenvalue $4{\bar \bW} \bW $.

${}$The off-diagonal elements of the mass matrix
$\cM_{\rm h}^2$ (\ref{hyper-mass})
for the hypermultiplet in the antisymmetric representation
vanish, since $\bQ^i \, \bQ_i = 0$.
There are $4(N-1)$ eigenvectors of $\cM_{\rm h}^2$
with the  eigenvalue $ {\bar \bW} \bW + \frac{N}{N-1}{\bar \bQ}^j \bQ_j$;
the remaining eigenvalues vanish.
The corresponding massive quantum
degrees of freedom can be parametrized in the form
\be
q = {1 \over \sqrt{2} }\,
\left(\begin{array}{cc}
A & B \\
C & A^{\, \rm T}
\end{array} \right),
\ee
where
\be
A = \left( \begin{array}{cl}
0 & ~\vec{a}_2^{\, \rm T}  \\
\vec{a}_1 & ~{\bf 0}_{N-1}
\end{array}\right)~, \quad B = \left( \begin{array}{cl}
0 & ~-  \vec{b}^{\, \rm T}  \\
\vec{b} & ~ {\bf 0}_{N-1}
\end{array}\right)~, \quad C = \left( \begin{array}{cl}
0 & ~ - \vec{c}^{\, \rm T}  \\
\vec{c} & ~{\bf 0}_{N-1}
\end{array}\right)~.
\ee
Again, $\vec{a}_1,$
$\vec{a}_2,$ $\vec{b}$ and $\vec{c}$ are  $(N-1)$-vectors.

Although some of the  hypermultiplet degrees of freedom in the
fundamental representation are
massive in the chosen background, they decouple.

The deformed conformal transformations (\ref{simpleh}) and
(\ref{simplea}) take the form
\bea
\d_{(1)} \bV_m &=&
-\frac{g^2}{4 \pi^2} \, (\partial^n \sigma) \,
\bF_{nm} \, \Big(
\frac{N-1}{ {\bar \bW} \bW + \frac{N}{N-1}{\bar \bQ}^j \bQ_j}
+\frac{1}{2 {\bar \bW} \bW} \Big) ~, \non \\
\d_{(1)} \bW &=&
-\frac{g^2}{4 \pi^2} \, (\partial^n \sigma) \,
(\partial_n \bW ) \, \Big(
\frac{N-1}{ {\bar \bW} \bW + \frac{N}{N-1}{\bar \bQ}^j \bQ_j}
+\frac{1}{2 {\bar \bW} \bW} \Big) ~; \label{vm-tra} \\
\d_{(1)} \bQ_i\ &=& -\frac{g^2}{4 \pi^2} \, (\partial^n \sigma) \,
(\pa_n \bQ_i)\, \frac{N} { {\bar \bW} \bW + \frac{N}{N-1}{\bar \bQ}^j \bQ_j} ~.
\label{hyp-tra}
\eea
The  vector multiplet
and the hypermultiplet transformations  differ due to
eigenvectors with eigenvalue $4{\bar \bW} \bW$, which are present in
$\cM_{\rm v}^2$  but absent in $\cM_{\rm h}^2.$

Unlike the situation  in the $\cN=4$ SYM theory,
the one-loop deformed transformations
$(\d_{\rm c}  +  \d_{(1)} )\,  \F$ specified in (\ref{vm-tra})
and (\ref{hyp-tra}) do not
realize the conformal algebra for finite values of $N$,
but only in a large $N$ limit. To realize the conformal
algebra for finite $N$,
it is necessary to include two and
higher-loop deformations to the conformal transformations of the fields;
these can be determined order by order from the
requirement of closure of the conformal algebra.
In particular, the two-loop deformation can be shown by this  means
to have the form
\bea
\d_{(2)} \bV_m &=& -{1 \over N} \,
\left( \frac{N g^2}{4 \pi^2} \right)^2
(\partial^n \sigma) \,\bF_{nm}\,
(\partial^p {\bar \bQ}^j)  (\pa_p \bQ_j) \;
\O (\bW, \bQ )
~,\non \\
\d_{(2)} \bW &=& -{1 \over N} \,
\left( \frac{N g^2}{4 \pi^2} \right)^2
(\partial^n \sigma)  (\pa_n \bW)
(\partial^p {\bar \bQ}^j)  (\pa_p \bQ_j) \;
\O (\bW, \bQ )
~, \\
\d_{(2)} \bQ_i\ &=& {1 \over N} \,
\left( \frac{N g^2}{4 \pi^2} \right)^2
(\partial^n \sigma)  (\pa_n \bQ_i)
(\partial^p {\bar \bW})  (\pa_p \bW) \;
\O (\bW, \bQ ) ~,
\non
\eea
where
\be
\O (\bW, \bQ ) =
\frac{1}{({\bar \bW} \bW + \frac{N}{N-1}{\bar \bQ}^j \bQ_j )^2}
\Big\{ \frac{1}{2 {\bar \bW} \bW }
- \frac{1}{ {\bar \bW} \bW + \frac{N}{N-1}{\bar \bQ}^j \bQ_j}
\Big\}~.
\ee

The field theory under consideration
possesses the global symmetries  $SU(2)_\cR \times U(1)_\cR
\times SO(8) \times SU(2)$ of which the first two factors
denote the $\cR$-symmetry.
The group $SO(8)$ rotates the four fundamental hypermultiplets,
while $SU(2)$ acts on the antisymmetric
hypermultiplet.\footnote{In the harmonic superspace approach \cite{GIOS},
the symmetries $SO(8)$ and $SU(2)$ can be realized
as (a combination of flavour and)
Pauli-G\"ursey transformations of $q^+$  hypermultiplets.}
Since $\cQ_{\rm F} = 0$, we cannot probe $SO(8)$
for the background chosen.
But the symmetries  $SU(2)_\cR \times U(1)_\cR \times SU(2)$
should be manifestly realized in the low energy effective action.
It is not difficult to see that both the one-loop
and two-loop conformal deformations respect these symmetries.

In the limit $N \to \infty$, the one-loop deformations
(\ref{vm-tra}) and (\ref{hyp-tra}) lead to conformal 
transformations of the form (\ref{n=4dct}), which are symmetries
of the Born-Infeld action (\ref{d3brane}). 
On symmetry grounds, when $N$ is finite, the low energy effective action of 
the field theory under consideration 
cannot have the Born-Infeld form considered in sects. 
4 and 5.

\sect{Kachru-Silverstein model}

Here we consider the simplest quiver gauge 
theory \cite{DM,JM,KS} --
$\cN=2$ super Yang-Mills theory
with gauge group $SU(N) \times SU(N)
\equiv SU(N)_L \times SU(N)_R $
and two hypermultiplets, $H_{i}$ and
$\tilde{H}_{i}$,
in the representations $(\bN, {\bar \bN} )$ and $({\bar \bN}, \bN)$
of the gauge group. Both $H_{i }$ and
$\tilde{H}_{i}$ carry an 
index of the automorphism group of the
$\cN=2$ supersymmetry algebra.
The hypermultiplets transform under
$SU(N)_L \times SU(N)_R $
as
\be
\d H = {\rm i}\, \z_L\, H - {\rm i} \, H \, \z_R~,
\qquad
\d \tilde{H} = {\rm i}\,  \z_R \, \tilde{H}
- {\rm i} \, \tilde{H}\, \z_L~,
\ee
where $ \z = \z_{a} t_{\au}$, with $t_{\au}$ the generators of $SU(N)$.
For simplicity we take the gauge coupling constants in the
two $SU(N)$ factors to be equal. 

Eq. (\ref{flat}) specifies the flat directions
in massless $\cN=2$ super Yang-Mills theories.
We are interested in those solutions of (\ref{flat})
in the Kachru-Silverstein model
which allow for non-vanishing hypermultiplet components.
Notationally, we now have $W= W_L \otimes {\bf 1}
+{\bf 1} \otimes W_R$ and $Q_i = (H_i, ~\tilde{H}_i)$.

Up to a gauge transformation, the general solution to the first
equation in (\ref{flat}), $[\cW, {\bar \cW}] =0$, is
$\cW = \cW_L \otimes {\bf 1}  + {\bf 1} \otimes \cW_R$,
with $\cW_L$ and  $\cW_R$ diagonal traceless matrices.
The second equation in (\ref{flat}), $\cW \cQ_i = 0$, becomes
\be
\cW_L\, \cH - \cH \, \cW_R =0~, \qquad
\cW_R \, \tilde{\cH}  - \tilde{\cH} \, \cW_L =0~.
\ee
These equations are obviously solved by any diagonal matrices
$\cW_L = \cW_R$, $\cH$ and $\tilde{\cH}$.
The third equation in (2.4),
${\bar \cQ}_{(i} T_{\au} \cQ_{j)} = 0$, is now
\bea
{\rm tr}\,\Big(\overline{\cH}_{(i} \, t_{\au} \,\cH{}_{j)}\Big)
-{\rm tr}\, \Big( \tilde{\cH}_{(i} \, t_{\au} \,
\overline{ \tilde{\cH} }_{j)} \Big) = 0~,
\quad
{\rm tr}\,\Big(\cH_{(i} \, t_{\au} \, \overline{\cH}{}_{j)}\Big)
-{\rm tr}\, \Big(
\overline{\tilde{\cH}}_{(i} \, t_{\au} \,
\tilde{\cH}_{j)} \Big) = 0~.
\label{ks-H}
\eea
The moduli space of vacua of the model thus includes
the  field configuration\footnote{The choice
$\overline{\tilde{\cH}}{}_{i} = \cH_i$, which
appears to solve (\ref{ks-H}) at first sight, is in fact not a solution.
The conjugation rules
$(Q_i)^\dagger = \overline{Q}^i  ~\Longleftrightarrow ~
(H_i)^\dagger = \overline{H}^i,~
(\tilde{H}_i)^\dagger = \overline{\tilde{H}}{}^i$
imply
$\left(\overline{\tilde{\cH}}_i \right)^\dagger = -\tilde{\cH}^i$,
and hence $\tilde{\cH}_i = - \overline{\cH}{}_i$.
}
\bea
\cW_L &=& \cW_R = {\bW \over N \sqrt{2  (N-1)}} \,
{\rm diag}(N-1, -1, \ldots , -1)~, 
\non \\
\cH_i &=& \tilde{\cH}{}_{i} =
{ \bQ_i \over \sqrt{2}} \,{\rm diag}(1,  0, \ldots , 0)~.
\eea
which preserves an unbroken gauge group
$SU(N-1) \times SU(N-1)$ together with the
diagonal $U(1)$ subgroup in $SU(N)_L \times SU(N)_R $
associated with the $\cW$ chosen.
In such a background,  the hypermultiplet mass matrix has $4(N-1)$ eigenvalues
$  \frac{1}{N-1}{\bar \bW} \bW + \frac{1}{N}{\bar \bQ}^j \bQ_j $ and 
one eigenvalue $\frac{4 (N-1)}{N^2} \,{\bar \bQ}^j \bQ_j.$
The massive degrees of freedom can be
parametrized
in the form
\be
 h_i = \left( \begin{array}{cl}
\a_i & ~ \, \vec{s}_i^{\, \rm T}  \\
\vec{r}_i & ~{\bf 0}_{N-1}
\end{array}\right)~, \quad \tilde{h}_i = \left( \begin{array}{cl}
\tilde{\a}_i &  ~\, \vec{\tilde{s}}_i^{\, \rm T}  \\
\vec{\tilde{r}}_i & ~{\bf 0}_{N-1}
\end{array}\right).
\ee
Here, all of the $(N-1)$-vectors $\vec{r}_i,$ $\vec{\tilde{r}}_i,$
$\vec{s}_i$
and  $\vec{\tilde{s}}_i$ are eigenvectors of $\cM_{\rm h}^2$
with the eigenvalue $ \frac{1}{N-1}{\bar \bW} \bW + 
\frac{1}{N} {\bar \bQ}^j \bQ_j,$
while the
eigenvector with eigenvalue $\frac{4 (N-1)}{N^2} \,{\bar \bQ}^j \bQ_j$
is $(\a_i - \tilde{\a}_i)/\sqrt{2}.$
This linear combination is orthogonal to the massless eigenvector
$(\a_i + \tilde{\a}_i)/ \sqrt{2}$ corresponding to an unbroken linear
combination of
$U(1)$ generators from $SU(N)_L$ and  $SU(N)_R.$

The mass matrix for the adjoint scalars and the gauge bosons also
 has $4(N-1)$ eigenvalues
$\frac{1}{N-1}{\bar \bW} \bW + \frac{1}{N} {\bar \bQ}^j \bQ_j$ and one eigenvalue
$\frac{4 (N-1)}{N^2} \,{\bar \bQ}^j \bQ_j.$ The adjoint  massive
degrees of freedom
can be parametrized in the form
\bea
w_L &=& {1 \over \sqrt{2N} }\,
\left(\begin{array}{cc}
\sqrt{\frac{2(N-1)}{N}} \, \a_L &  \, \vec{r}_L^{\, \rm T} - {\rm i}  
\vec{s}_L^{\, \rm T} \\
\vec{r}_L + {\rm i} \vec{s}_L & - \sqrt{\frac{2}{N(N-1)}} \, \a_L \, 
{\bf 1}_{N-1}
\end{array} \right)~, \non \\
 w_R &=&
{1 \over \sqrt{2N} }\, \left(\begin{array}{cc}
\sqrt{\frac{2(N-1)}{N}} \, \a_R &  \, \vec{r}_R^{\, \rm T} - {\rm i} 
 \vec{s}_R^{\, \rm T} \\
\vec{r}_R + {\rm i} \vec{s}_R & - \sqrt{\frac{2}{N(N-1)}} \, \a_R \, 
{\bf 1}_{N-1}
\end{array} \right).
\eea
The $(N-1)$-vectors $\vec{r}_L,$ $\vec{s}_R,$  $\vec{r}_L$
and  $\vec{s}_R$ are eigenvectors of $\cM_{\rm v}^2$
with the eigenvalue $\frac{1}{N-1}{\bar \bW} \bW + \frac{1}{N}{\bar \bQ}^j \bQ_j,$
while the
eigenvector with eigenvalue $\frac{4(N-1)}{N^2} \,{\bar \bQ}^j \bQ_j$
is $(\a_L - \a_R)/\sqrt{2}.$

The scalars (both adjoint and hypermultiplet)
 with mass $\frac{1}{N-1}{\bar \bW} \bW + \frac{1}{N}{\bar \bQ}^j \bQ_j $
couple to the gauge bosons of the same mass via a derivative of the
corresponding background scalar. However,  for the adjoint scalars, the linear
combination  $(\a_L - \a_R)/\sqrt{2}$  does not couple to the
gauge bosons, and so there is no contribution from this mass
eigenstate  to the quantum
corrected transformations. This is to be contrasted with
the hypermultiplet mass eigenstate $(\a_i - \tilde{\a}_i)/\sqrt{2}$ with
the mass
 $\frac{4 (N-1)}{N^2} \,{\bar \bQ}^j \bQ_j$,
which  couples to the gauge boson with mass
$\frac{4 (N-1)}{N^2} \,{\bar \bQ}^j \bQ_j,$
giving rise to an additional term in the one-loop hypermultiplet
transformation compared with the adjoint scalar transformation.  This is the
``reverse'' of the situation encountered in the $USp(2N)$ example.
The one-loop conformal deformations (\ref{vector1}) and
(\ref{scalar1}) are
\bea
\d_{(1)} \bV_m &=&
-\frac{g^2}{4 \pi^2} \, (\partial^n \sigma) \,
\bF_{nm} \,
\frac{N}{ \frac{N}{N-1}{\bar \bW} \bW + {\bar \bQ}^j \bQ_j}  ~, \non \\
\d_{(1)} \bW &=&
-\frac{g^2}{4 \pi^2} \, (\partial^n \sigma) \,
(\partial_n \bW ) \,
\frac{N}{ \frac{N}{N-1}{\bar \bW} \bW + {\bar \bQ}^j \bQ_j}  ~;
\label{ks-vm-tra} \\
\d_{(1)} \bQ_i\ &=& -\frac{g^2}{4 \pi^2} \, (\partial^n \sigma) \,
(\pa_n \bQ_i)\, \Big(
\frac{N-1} {\frac{N}{N-1} {\bar \bW} \bW + {\bar \bQ}^j \bQ_j}
+ \frac{1}{4 {\bar \bQ}^j \bQ_j } \Big)~.
\label{ks-hyp-tra}
\eea

Again, the one-loop deformed transformations
$(\d_{\rm c}  +  \d_{(1)} )\,  \F$ specified by (\ref{ks-vm-tra})
and (\ref {ks-hyp-tra})
only provide a realization of the
conformal algebra in a large $N$ limit. For finite values of $N,$ it
is necessary to include higher-loop corrections, which can again be
determined order by order in $g^2 N$
by requiring closure of the algebra of conformal
transformations of the fields. Using this procedure, the two-loop
conformal deformation is found to be
\bea
\d_{(2)} \bV_m &=&  {1 \over N} \,
\left( \frac{N g^2}{4 \pi^2} \right)^2
(\partial^n \sigma) \, \bF_{nm}\,
(\partial^p {\bar \bQ}^j)  (\pa_p \bQ_j) \;
\O (\bW, \bQ )
~,\non \\
\d_{(2)} \bW &=& {1  \over N} \,
\left( \frac{N g^2}{4 \pi^2} \right)^2
(\partial^n \sigma)  (\pa_n W)
(\partial^p {\bar \bQ}^j)  (\pa_p \bQ_j) \;
\O (\bW, \bQ )
~, \\
\d_{(2)} \bQ_i\ &=& -{1 \over N} \,
\left( \frac{N g^2}{4 \pi^2} \right)^2
(\partial^n \sigma)  (\pa_n \bQ_i)
(\partial^p {\bar \bW})  (\pa_p \bW) \;
\O (\bW, \bQ )
~,
\non
\eea
where
\be
\O (\bW, \bQ ) =
\frac{1}{(\frac{N}{N-1}{\bar \bW} \bW + {\bar \bQ}^j \bQ_j )^2}
\Big\{ \frac{1}{4 {\bar \bQ}^j \bQ_j }
- \frac{1}{ \frac{N}{N-1}{\bar \bW} \bW + {\bar \bQ}^j \bQ_j}
\Big\}~.
\ee

The field theory under consideration
possesses the global symmetries  $SU(2)_\cR \times U(1)_\cR
\times SU(2)$, where the group $SU(2)$ mixes the hypermultiplets
$H$ and $\overline{\tilde{H}}$. 
It is not difficult to see that both the one-loop
and two-loop conformal deformations respect these symmetries.

\vskip.5cm

\noindent
{\bf Acknowledgements.}
The work of ST is supported in part by GIF, the German-Israeli Foundation
for Scientific Research. He would like to thank the Theory Division
of CERN where this work was completed and A. Armoni and A. Font for valuable
discussions. 
The work of SMK and INM is partially supported by 
a University of Western Australia Small Grant and an ARC Discovery Grant.
SMK is thankful for kind hospitality extended to him 
at the Max Planck Institute for Gravitational Physics 
(Albert Einstein Institute), Golm where this project was commenced.
His work was also supported in part by the Alexander 
von Humboldt Foundation and the Australian Academy of 
Science.

\end{document}